\documentclass[reprint]{revtex4-1}
\setcitestyle{square}
\usepackage[utf8]{inputenc}
\usepackage{graphicx,url, hyperref, microtype, subcaption, graphicx, dcolumn, bm, dsfont, xcolor, csquotes, dsfont, amsmath, amsbsy}
\usepackage[utf8]{inputenc}
\usepackage[T1]{fontenc}
\usepackage[margin=1in]{geometry}
\usepackage[sort&compress]{natbib}

\newcommand{\basis}[1]{\hat{\boldsymbol{e}}_{#1}}
\newcommand{\tbasis}[2]{\basis{#1}\otimes\basis{#2}}
\newcommand{\fbasis}[4]{\tbasis{#1}{#2}\otimes\tbasis{#3}{#4}}
\newcommand{\ft}[1]{\tilde{\boldsymbol{#1}}}
\newcommand{\bs}[1]{{\boldsymbol{#1}}}

\newcommand{\Cdot}{\! \cdot \!}

\begin{document}
\title{{Poroelasticity as a Model of Soft Tissue Structure: \protect\\ Hydraulic Permeability Inference \\ for Magnetic Resonance Elastography in Silico}}
\author{Damian Sowinski}
\email{Damian.Sowinski@Dartmouth.edu}
\affiliation{Thayer School of Engineering, Dartmouth College, Hanover, NH 03755, USA}
\author{Matthew McGarry}
\affiliation{Thayer School of Engineering, Dartmouth College, Hanover, NH 03755, USA}
\author{Scott Gordon-Wylie}
\affiliation{Thayer School of Engineering, Dartmouth College, Hanover, NH 03755, USA}
\author{Elijah Van Houten}
\affiliation{D\'epartement de g\'enie m\'ecanique, Universite de Sherbrooke, Sherbrooke, Quebec, Canada}
\author{John Weaver}
\affiliation{Thayer School of Engineering, Dartmouth College, Hanover, NH 03755, USA}
\affiliation{Department of Radiology, Dartmouth-Hitchcock Medical Center, Lebanon, NH 03756 USA}
\affiliation{Geisel School of Medicine, Dartmouth College, Hanover, NH 03755, USA}
\author{Keith Paulsen}
\affiliation{Thayer School of Engineering, Dartmouth College, Hanover, NH 03755, USA}
\affiliation{Geisel School of Medicine, Dartmouth College, Hanover, NH 03755, USA}
\affiliation{The Center for Surgical Innovation, Dartmouth-Hitchcock Medical Center, Lebanon, NH 03756 USA.}
\date{\today}
\begin{abstract}
Magnetic Resonance Elastography allows noninvasive visualization of tissue mechanical properties by measuring the displacements resulting from applied stresses, and fitting a mechanical model.
Poroelasticity naturally lends itself  to describing tissue - a biphasic medium, consisting of both solid and fluid components. 
This article reviews the theory of poroelasticity, and shows that the spatial distribution of hydraulic permeability, the ease with which the solid matrix permits the flow of fluid under a pressure gradient, can be faithfully reconstructed without spatial priors in simulated environments. 
The paper describes an in-house MRE computational platform - a multi-mesh, finite element poroelastic solver coupled to an artificial epistemic agent capable of running Bayesian inference to reconstruct inhomogenous model mechanical property images from measured displacement fields. 
Building on prior work, the domain of convergence for inference is explored, showing that hydraulic permeabilities over several orders of magnitude can be reconstructed given very little prior knowledge of the true spatial distribution.
\tiny
{\fontsize{8}{11}Elastography, MRI, Biophysics, Biomechanics, Poroelasticity, Simulation, Bayesian Inference, Tissue Structure, Brain, Epistemology, Continuum Mechanics}
\end{abstract}
\maketitle
\tableofcontents
\newpage
\section{Introduction}%
In the past two decades magnetic resonance elastography (MRE) has emerged as an imaging modality that extends the capabilities of standard MRI, giving clinicians and researchers unprecedented access to tissue properties in vivo.
As it develops rapidly, MRE promises clinicians information akin to the data surgeons acquire through manual palpation - material properties such as stiffness - but noninvasively and quantitatively, with applications in diagnosis and monitoring of diseases.
Connecting macroscopic mechanical response to microscopic cellular organization of tissue, MRE opens up new avenues for assessing the etiology of tissue pathology.
The modality has been implemented to varying degrees of success on skeletal muscle \cite{Ehman:2001:a,Sack:2002:a,Ehman:2006:b,Ehman:2007:b,KaiNan:2008:a,Sack:2010:a,Podrekar:2019:a, Gao:2018:a,Sinkus:2019:a}, breast \cite{McKnight:2002:a, Sinkus:2002:a,Paulsen:2003:a,Sinkus:2005:a,Sack:2008:a,Kolipaka:2017:a,Sinkus:2018:a, Sinkus:2018:b}, liver \cite{Ehman:2006:a,Ehman:2007:a,Araki:2010:a,Ehman:2010:a,Ehman:2013:a,Johnson:2018:a,Kundapor:2018:a}, and brain\cite{Gao:2007:a,Ehman:2008:a}.

Elastography dates back to Gierke et al's 1952 paper, {\it Physics of Vibrations in Living Tissue} - one can discern tissue properties by observing a tissue's response to vibration\cite{Gierke:1952:a}.
MRE is multidisciplinary at its core, requiring expertise in magnetic resonance imaging, biophysical modelling/theoretical mechanics, and numerical simulations/machine learning.
Several reviews are available \cite{Manduca:2001:a,Mariappan:2010:a,Sarvazyan:2011:a,Tang:2015:a,Low:2016:a,Bayly:2018:a}. 

The purpose of the current paper is to explore a poroelastic model of tissue, and convey novel results on the simultaneous imaging of both shear stiffness and hydraulic permeability in silico.
These results extend prior studies on poroelastic parameter estimation\cite{Perrinez:2007:a,Perrinez:2010:a,Pattison:2014:a,Tan:2017:a}, and quantify the domain of valid inference.
We show that convergence to the correct in-silico property values is possible even when prior knowledge is off by several orders of magnitude.
These results are a foundation for ongoing in vitro and in vivo work, giving reasonable bounds for noiseless data.

This paper is organized as follows:
Section \ref{sec:2} covers the theoretical details of our tissue model and how inference is accomplished. 
It also details some of the numerical considerations that have been implemented in our in-house MRE computational platform.
Section \ref{sec:3} details the setup of our simulations - inference of one and two properties from in-silico data generated by solving Newton's Laws for a poroelastic medium.
Single and dual property contrast conditions necessary for fidelitous inference are examined.
In Section \ref{sec:4} we discuss the results from our simulations and conclude with planned future directions for in-vitro and in-vivo application.\\

\subsection*{Notational Conventions}

Throughout, we use coordinate independent tensor notation to emphasize the physical underpinnings of expressions, but, when required to use coordinates, follow the {\it Einstein summation convention} where repeated indices in terms are summed over. 
Indices are drawn from the middle of the Latin alphabet, $i,j,k,\dots$, and run from $1$ to $3$.
Scalar quantities are written in standard font with no typographical emphasis.
Vectors and second rank tensors appear in lower case bolded font, the former in Latin and latter in Greek.
Fourth rank tensors are bolded, and in capitalized Latin.
When discussing either solid or fluid phases similar symbols are used to denote physical quantities, however the addition of a subscript $s$ or $f$, respectively, is added to ensure physical clarity.

We denote the standard tensor product with a $\otimes$.
An orthonormal vector basis is denoted $\{\basis{i}\}_{i=1}^3$, with $\basis{i}\Cdot\basis{j}=\delta_{ij}$ where the RHS is the Kronecker delta.
At times we use the Cartesian representation of the basis, $\basis{1}=\hat{\boldsymbol{x}},\basis{2}=\hat{\boldsymbol{y}}$, and $\basis{3}=\hat{\boldsymbol{z}}$.
We construct the basis for second and fourth rank tensors from these, $\{\basis{ij}=\tbasis{i}{j}\}_{i,j=1}^3$ and $\{\basis{ijk\ell}=\fbasis{i}{j}{k}{\ell}\}_{i,j,k,\ell=1}^3$ respectively.
The second rank identity tensor is denoted $\mathds{1}=\basis{ii}$.
The dot product, $\Cdot$ , and double dot product, $:$ , act on neighboring basis tensors via $\basis{i}\Cdot\basis{j}=\delta_{ij}$ and $\basis{ij}:\basis{k\ell}=\delta_{ik}\delta_{j\ell}$, implying contraction on the component indices.
For example, given an arbitrary vector, $\bs{a}$, second rank tensor, $\bs{\beta}$, and fourth rank tensor, $\bs{C}$, we can construct a vector as follows:
\begin{widetext}
\begin{align}
    \bs{a}\Cdot\bs{C}:\bs{\beta}&=(a_{i}\basis{i})\Cdot(C_{jk\ell m}\basis{jk\ell m}):(\beta_{no}\basis{no})\nonumber\\
    &=a_{i}C_{jk\ell m}\mathit{\beta}_{no}\ \basis{i}\Cdot\fbasis{j}{k}{\ell}{m}:\tbasis{n}{o}\nonumber\\
    &=a_{i}C_{jk\ell m}\mathit{\beta}_{no}\ \delta_{ij}\basis{k}\delta_{\ell n}\delta_{mo}\nonumber\\
    &=a_iC_{ik\ell m}\beta_{\ell m}\basis{k}\nonumber
\end{align}
\end{widetext}
The index gymnastics in evaluating such expressions can get pretty messy, which is why we stick to coordinate independent notation as much as possible.
The transpose acts on second rank basis tensors as $\basis{ij}^T=\basis{ji}$.
The gradient operator is symbolized by a nabla, $\nabla$.
Temporal Fourier transforms of quantities appear with the same typographical emphasis as above, having the added decor of a tilde.
For temporal partial derivatives we employ fluxion notation, so that $\dot f \equiv \frac{\partial f}{\partial t}$, $\ddot f \equiv \frac{\partial^2\!\!f}{\partial t^2}$, etc. 

\section{Modelling Tissue}\label{sec:2}

Biological tissue is a complex medium, the result of almost four billion years of continual evolutionary pressure and the whims of chance.
At the microscopic level, the variety of cellular structures in animals would require a vastly more complicated set of modelling assumptions than continuum mechanics can provide. 
However, tracing over these microscopic degrees of freedom, an effective theory of tissue should have more in common with a coupled solid-fluid description than one consisting of a single continuum. 
Such a model already exists, developed extensively during the last century by the geophysics community - poroelasticity.\\

\subsection*{Poroelasticity}

\begin{figure}[p]  
    \centering
    \includegraphics[width=4.00in]{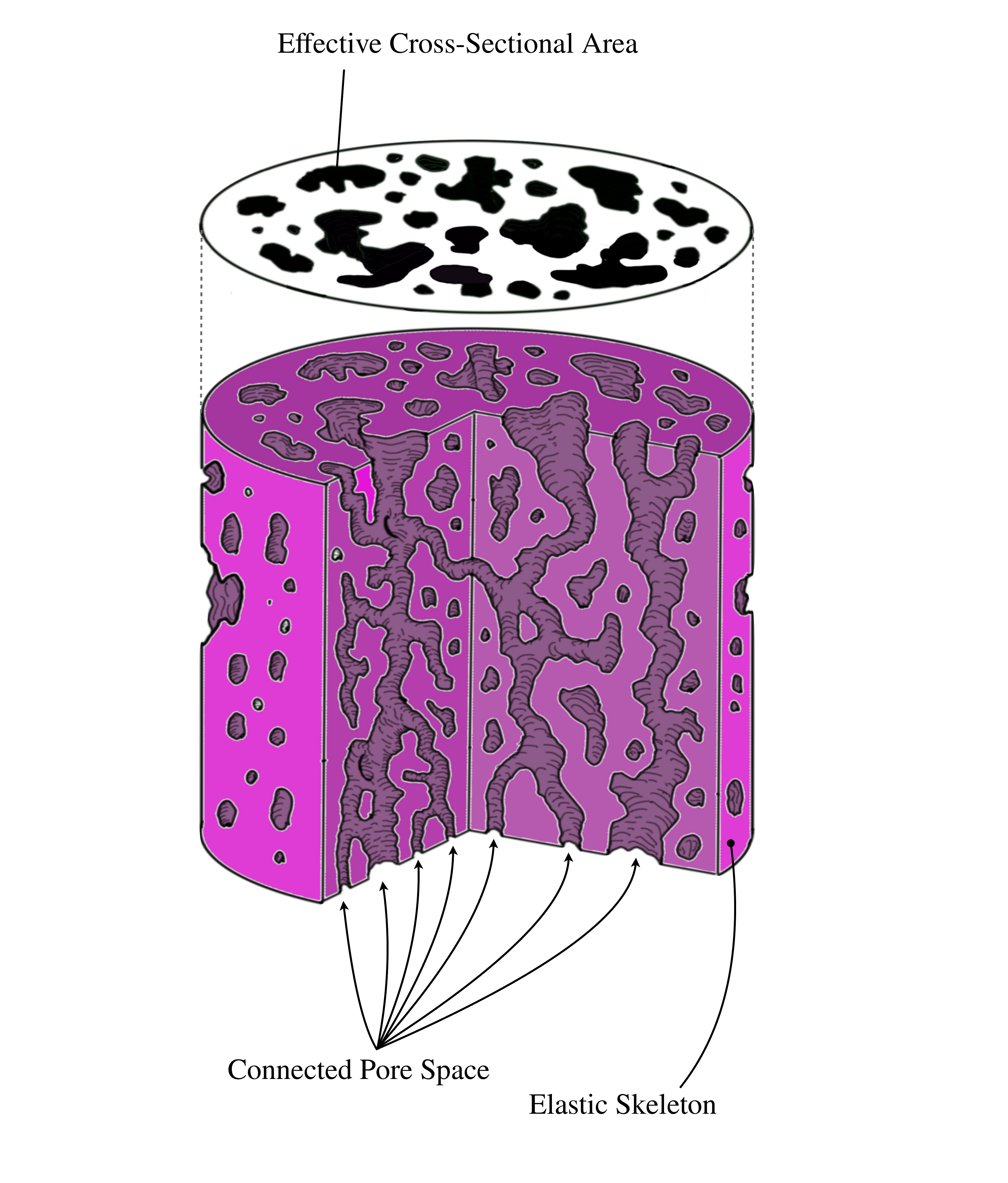}
    \caption{
        The model of a porous material consists of an elastic matrix - the solid - and a connected pore space. 
        Pores whose connected boundaries are homeomorphic (topologically identical) to spheres are not a part of the connected pore space, as nothing can flow through them.
        The connected pore space is filled with a fluid, whose interactions with the walls of the elastic skeleton provide a much richer phenomenology than elastic/viscoelastic models of materials.
        When a porous substance is intersected with a plane, the cross section, whose area is $A$, will have holes in it.
        The effective cross-sectional area, $A_f$, is the area of all of these holes. 
        For an isotropic material, the orientation of the intersecting plane does not change the ratio $A_f/A$.
        The Delesse-Rosiwal law states that this ratio is equal to the ratio of the pore space volume to the total volume i.e. the porosity \cite{Delesse:1848:a}.
        }
        \label{fig:Porous Material}
\end{figure}

Poroelasticity is an effective mechanical model describing the continuum as a coarse grained biphasic medium consisting of an {\it elastic matrix} coupled to a {\it fluid}.
The former is not simply-connected, but has a connected pore space with a tortuous topology - it is this pore space that the fluid moves through.
The coupled dynamics of solid and fluid result in a much richer phenomenology than single phase continuum models.
Fig.\ref{fig:Porous Material} shows a schematic of a porous structure.

\begin{figure}[p]  
    \centering
    \includegraphics[width=4.00in]{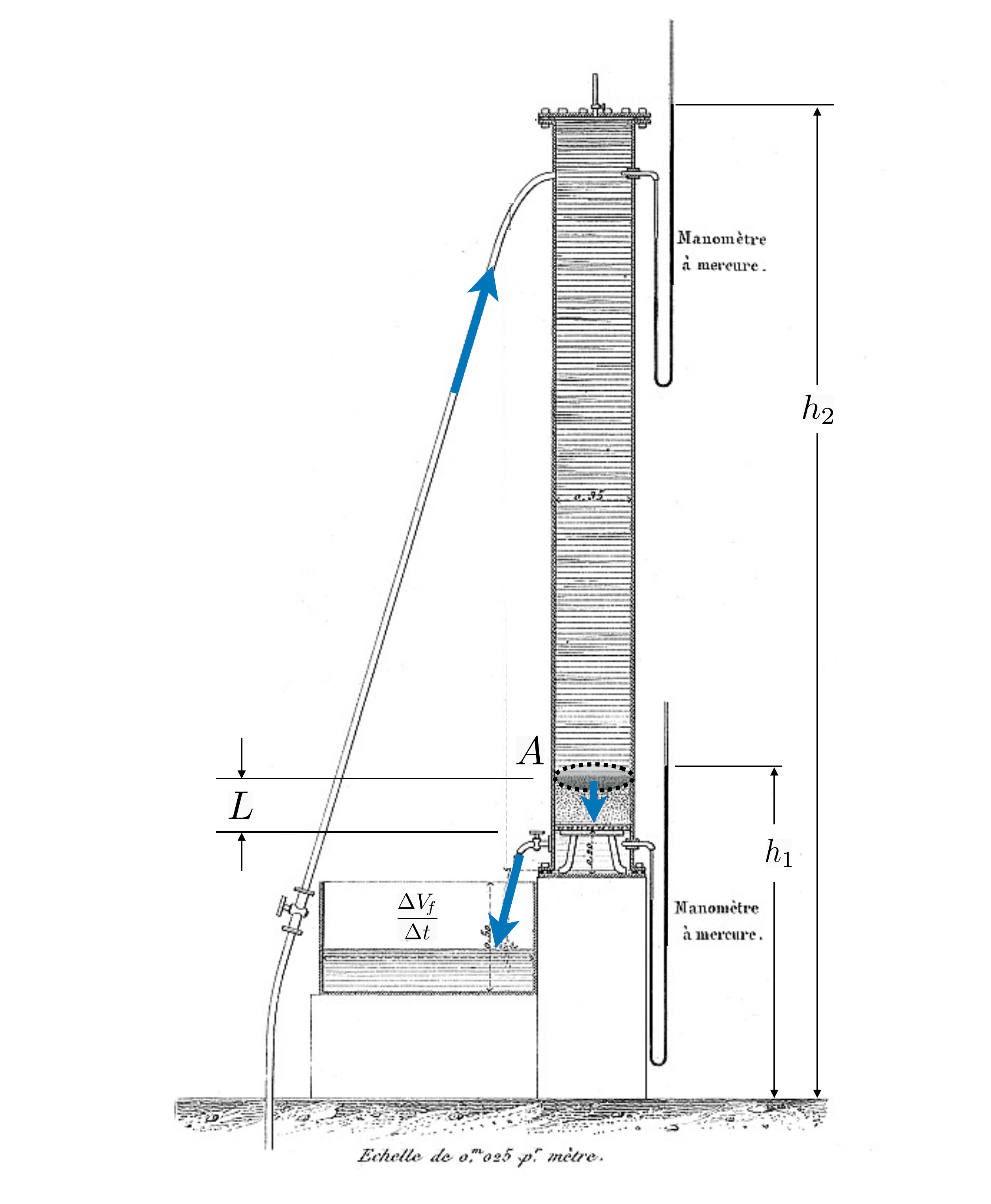}
    \caption{
        The setup of one of the sand column experiments, from plate 24, fig.3 of Darcy's 1856 {\it Les Fontaines Publiques de la Ville de Dijon} \cite{Darcy:1856:a}. 
        Annotations have been added to clarify the the origin of the physical quantities appearing in Eq.(\ref{eq:Darcy's Law}).
        The piezometric heads $h_1$ and $h_2$ in the two piezometers on the right are measured using mercury.
        $L$ is the length of the sand column, and $A$ its cross-sectional area.
        In this setup the diameter of the column is $0.35$m, and the height of the column is $3.5$m. 
        $\Delta V_f/\Delta t$ is the rate at which fluid volume is leaving the column, measured by observing the rise of the water level in the container.
        One can quickly see that this particular setup is problematic for applying Eq.(\ref{eq:Darcy's Law}) directly unless the sand column reaches up to near the top piezometer.
        If not, then the piezometric head must be replaced by the total head so as to includes the pressure head coming from the column of water.
        }
        \label{fig:Darcy Experiment}
\end{figure}

The theory of poroelasticity has its roots in the empirical work of Henry Philibert Gaspard Darcy, a water engineer tasked in 1856 with evaluating the many elaborate public fountains within the city of Dijon, France.
Darcy, who had been working as a water engineer for nearly three decades, was pivoting towards pure research, and used this as an opportunity to to write his $600$ page magnum opus {\it Les Fontaines Publiques de la Ville de Dijon}, wherein he describes the sand column experiments pictured in Fig.\ref{fig:Darcy Experiment} \cite{Darcy:1856:a}.
By examining the flow of water through sand in cylindrical pipes, Darcy discovered a linear proportionality between the rate of fluid volume leaving the bottom of a pipe, $\Delta V_f/\Delta t$, and the difference in piezometric head at the ends of the pipe, $\Delta h$, 
\begin{equation}\label{eq:Darcy's Law}
    \frac{1}{A}\frac{\Delta V_f}{\Delta t}=K\frac{\Delta h}{L},
\end{equation}
where $L$ is the length of the pipe, and $A$ its cross-sectional area.
{\it Hydraulic conductivity} is the name given to the constant of proportionality, $K$, and for water flowing through sand was measured to range from $2.07-3.22\!\times\! 10^{-4} \text{m}$/$\text{s}$\cite{Brown:2002:a}.
This name is kept alive today by geologists, though we will see shortly that other fields' naming conventions have created an aura of ambiguity.
Note that $K$ depends on the properties of the fluid used in the piezometers, as well as the porous medium and the fluid flowing through said medium - since geology is primarily concerned with the filtration of water through soils, rocks, and clays, they think of it as intrinsic to the medium.
For us, biologically relevant fluids can differ substantially in density and viscosity, so we mustn't fall into the same mindset.
The empirical formula Eq.(\ref{eq:Darcy's Law}) came to be known as {\it Darcy's Law}, a starting point for understanding how the flow of a fluid is affected by its coupling to a porous material.

A modern view of Eq.(\ref{eq:Darcy's Law}) utilizes the instantaneous rate of fluid volume moving through a cylinder at constant velocity, $v_f$, replacing $\Delta V_f/\Delta t\rightarrow v_fA_f$, where $A_f$ is the effective area seen by the fluid due to the microscopic geometry of the porous medium.
The effective area is related to the {\it porosity} (volume fraction of fluid, assuming saturation) by the Delesse-Rosiwal law (second equality)\cite{Delesse:1848:a}, $\phi = V_f/V=A_f/A$. 
Identifying the pressure differential between the ends of the pipe with the drop in head via $\Delta P=\rho_f g\Delta h$ ( $\rho_f$ is the fluid density and $g$ the gravitational acceleration), shrinking the pipe to infinitesimal length, $L\rightarrow 0$, and restoring the tensorial nature of the physical quantities, we find
\begin{equation}\label{eq:discharge}
    \bs{q} = -\frac{1}{\eta}\bs{\kappa}\Cdot\nabla P
\end{equation}
where two clarifications are in order.
(1) The LHS is the {\it Darcy velocity} (or {\it filtration velocity}), $\bs{q}=\phi \bs{v}_f$, the magnitude of which is called the {\it specific discharge}.
(2)  The RHS is rewritten using the {\it hydraulic permeability tensor}, $\boldsymbol{\kappa}$ - a property belonging solely to the porous medium - and $\eta$ - the dynamic viscosity of the fluid.
The full justification of this separation is possible through a microscopic treatment using homogenization theory\cite{Coussy:2004:a,Lopatnikov:2004:a,Bear:2010:a,Cheng:2016:a}, extending the validity of Darcy's Law to pressure gradients not necessarily generated by gravitational forces.

For an isotropic medium, such as Darcy's sand column, $\bs{\kappa}=k\mathds{1}$, and we call $k$ the {\it hydraulic permeability}.
Unfortunately, as mentioned earlier, there is a bit of ambiguity within the literature depending on the scientific field one is enamoured with - geologists measure piezometric head and refer to Darcy's original $K$ [LT$^{-1}$] as hydraulic conductivity, while most other fields measure pressure gradients and refer to $\bar K=K/\rho_f g$ [L$^3$TM$^{-1}$] as such.
Furthermore, the medical and bio-engineering literature has, on occasion, used the terms hydraulic conductivity and permeability interchangeably; it is best to always check units to make sure that you and the author are on the same page.
Given the dynamic viscosity of water is  $\sim 10^{-3}\text{Pa}\! \cdot\! \text{s}$, this implies Darcy's original measurements of sand's hydraulic permeability on the order of $\sim 10^{-11}\text{m}^2$.
In honor of this, hydraulic permeability is now measured in units of Darcys, D, so that clean sand in aquifers comes out as $\sim10^1$D.
The unit is defined so that $1$D permits a filtration velocity of a cm$/$s for a fluid with dynamic viscosity $10^{-3}$Pa$\cdot$s under a pressure gradient of 1 atm$\!/\!$cm, or in SI, $1\text{D} =9.86923\!\times\! 10^{-13}\text{m}^2$.
The strangeness of this value comes from there being $101,325$Pa in $1$atm, rather than a round $10^5$, but who are we to quibble with history?

Karl von Terzaghi, a mechanical engineer,  laid out the dynamics of structure consolidation on soil in a series of papers\cite{Tergazhi:1922:a,Tergazhi:1925:a,Tergazhi:1943:a} between 1922-1925, extending Darcy's insights to become the {\it father of soil mechanics}.
He incorporated the dynamics of soil in response to the fluid by introducing an effective stress tensor, $\bs{\sigma}$, to describe the bulk medium,
\begin{equation}\label{eq:Terzaghi Decomposition}
    \bs{\sigma}= \bs{\sigma}_s - P\mathds{1}.
\end{equation}
Here $\bs{\sigma}_s$ is the stress tensor of the solid, and $P$ is the {\it pore pressure}.
The effective stress describes the net stress on the solid skeleton, and it's definition is referred to as {\it Terzaghi's Principle} - the total stress on the elastic matrix is lessened by the pore pressure.
Terzaghi's analysis describes a static, highly symmetric, and fully saturated system in which an incompressible fluid is confined to a one dimensional column of incompressible solid grains.
Rearrangements of these grains drive deformation of this {\it soil} when a load is applied.
The decomposition of the stress into an effective bulk and fluid is a bit like looking at the soil through poor lenses that smear the distinction between microscopic grains and pore fluid - a technique central to homogenization theory in engineering, and coarse graining in physics and chemistry\cite{Kadanoff:1966a,Wilson:1971a,Marrink:2007a}.
One should be a tad careful in their mind's eye in all that follows - in going to this effective picture we now imagine solid and fluid components existing simultaneously at every spatial point, their dynamics coupled as a result of coarse graining the microscopic structure.

Though revolutionary in application, Terzaghi's assumptions were ill suited for generality.
Thermodynamically speaking, his model contains the pair of dynamical variables $\{\bs{\sigma},P\}$, the effective stress and the pore pressure, respectively; the elastic matrix strain, \begin{equation}
 \bs{\varepsilon}=\frac{1}{2}\left(\nabla\!\otimes\!\bs{u}_s+(\nabla\!\otimes\!\bs{u}_s)^T\right),
\end{equation} 
is conjugate to the former, where $\bs{u}_s$ is the displacement field of the solid matrix.
A good guess for the  kinematic variable corresponding to the conjugate of the pressure would seemingly be the fluid strain rate, which for a perfect fluid with velocity field $\bs{v}_f=\dot{\bs{u}}_f$ is the single degree of freedom $\text{tr}(\dot{\bs{\varepsilon}}_f)=\nabla\Cdot\bs{v}_f$.
Though this works to describe a fluid alone and results in the Navier-Stokes equations, this term cannot be coupled to the strain in a linear theory, nor to the solid displacement as that ruins translational invariance.
The only option to preserve these necessary ingredients would be the fluid strain, $\nabla\Cdot\bs{u}_f$.
It turns out that this choice almost works; unfortunately it couples the solid and fluid a tad too weakly by predicting a vanishing pore pressure when the fluid moves in sync with the solid.
With the obvious choices for a kinematic conjugate used up, it appears as if we must continue our search in darkness.

Maurice Antony Biot, an applied physicist, shed light on the correct coupling in 1941 by introducing\cite{Biot:1941:a} the {\it variation in relative fluid content},
\begin{equation}
\zeta\!=-\! \nabla\Cdot\phi({\boldsymbol u}_f\!-\!{\boldsymbol u}_s)+\Gamma, 
\end{equation}
where $\gamma=\dot\Gamma$ is referred to as a {\it fluid source}.
His treatment generalized the Darcy velocity to a relative motion of the fluid with respect to the solid, $\bs{q}=\phi(\bs{v}_f-\bs{v}_s)$, thus allowing for an analysis of situations in which both the fluid and porous medium are in motion.
Note that the relative fluid content decreases when net flow out of a dilating volume of the solid matrix occurs, and increases if the flow is reversed - together with the filtration velocity this is summarized by a continuity equation
\begin{equation}\label{eq:continuity equation}
    \dot\zeta + \nabla\Cdot\bs{q}=\gamma.
\end{equation}
Here, we see the fluid source, $\gamma=\dot\Gamma$, in action either creating fluid content, $\gamma>0$, or annihilating it, $\gamma<0$ - a rather useful quantity when considering tissue near veins and arteries. 

In Terzaghi's analysis, bulk volume change is driven by rearrangements of the pore space and flow of the fluid.
Biot realized that this physics may be true for nearly incompressible sand grains, but for a softer elastic medium one should also take into account the contribution of solid compressibility to the bulk volume change.
An applied isotropic load can now change the volume of the bulk by altering the volume of the solid skeleton due to the fluid pushing back - for reasonably small volume changes the effective support of the fluid with regards to the bulk is then less than the fluid pressure. 
Biot argued that the correct effective stress should be a modification of Eq.(\ref{eq:Terzaghi Decomposition}), 
\begin{equation}\label{eq:Biot decomposition}
    \bs{\sigma}=\bs{\sigma}_s-\alpha P \mathds{1}
\end{equation}
with the {\it Biot effective stress coefficient}, $\alpha$, a function of the drained effective bulk modulus and the unjacketed solid bulk modulus, $K$ and $K'_s$ respectively\cite{Detournay:1993:a},
\begin{align}\label{eq:Biot coefficient}
    \alpha = 1-\frac{K}{K'_s}.
\end{align}
To clarify, experimentally the unjacketed bulk modulus is measured in undrained conditions wherein a sample is submerged in fluid; the fluid pressure is then allowed to vary and the volume change of the sample is measured. 
A jacketed test, on the other hand, is done under drained conditions wherein a dry sample is wrapped in a membrane and connected to the atmosphere via a small hole; a load is then applied under constant atmospheric pressure, and the volume change measured.
In aerated soil $K/K'_s\ll 1$, validating Terzaghi's treatment\cite{Skempton:1954:a,Bishop:1973:a}.
The theory predicts a second type of longitudinal wave within a porous medium, which was observed in 1979\cite{Plona:1980:a}. 
Biot refined his model by incorporating anisotropy, analyzing dispersion in wave propagation, and examining non-linear extensions of the constitutive relations, earning him the title of {\it father of poroelasticity} \cite{Biot:1955:a,Biot:1956:a,Biot:1956:b,Biot:1973:a}.

Examining the effective potential can further illuminate the role of $\alpha$ - it is the dimensionless coupling between the volumetric strain of the skeleton and the variation in relative fluid content.
Since it and $\zeta$ are both dimensionless, the {\it Biot modulus}, $M$, must be introduced to describe the energy density coming from the relative fluid content and its coupling to the elastic matrix, allowing the quadratic potential energy density for a linear poroelastic medium to be written as
\begin{align}\label{eq:thermodynamic potential}
    U &=\frac{1}{2}\bs{\varepsilon}:\bs{E}:\bs{\varepsilon}+\frac{1}{2}M\zeta^2-\alpha M \zeta \text{tr}(\bs{\varepsilon})
\end{align}
where $\bs{E}=\bs{E}_0+\alpha^2M\mathds{1}\!\otimes\!\mathds{1}$ is an effective elastic modulus, a fourth rank tensor describing the effective coupling between the components of the strain.
For an isotropic Hookean material described by Eq.(\ref{eq:Hookean material}), note that the Biot effective stress coefficient and modulus serve to renormalize the second Lam\'e paramter, $\lambda$.
The conjugate variables can now be computed from this, giving the {\it poroelastic constitutive relations}
\begin{align}\label{eq:constitutive relation 1}
    \bs{\sigma} &= \frac{\partial U}{\partial \bs{\varepsilon}}= \bs{E}:\bs{\varepsilon}-\alpha M\zeta \mathds{1}\\
    \label{eq:constitutive relation 2}
    P &=\frac{\partial U}{\partial \zeta}= -\alpha M \text{tr}(\bs{\varepsilon})+M\zeta
\end{align}
Combining Eq.(\ref{eq:constitutive relation 1}) and Eq.(\ref{eq:constitutive relation 2}) recovers the Biot decomposition, Eq.(\ref{eq:Biot decomposition}).
These expressions clarify the interpretation of $M$ - it is the change in pore pressure caused by a variation in water content under the constraint of no solid dilation, $M=\partial P/\partial \zeta|_{\text{tr}(\bs{\varepsilon})=0}$.
It can be written in terms of the Biot effective stress coefficient, as well as the fluid and jacketed solid moduli, $K_f$ and $K''_s$ respectively\cite{Detournay:1993:a,Cheng:2016:a},
\begin{align}\label{eq:Biot modulus}
    \frac{1}{M}=\frac{1}{K}\left(\alpha(1-\alpha)-\phi\frac{K}{K''_s}+\phi\frac{K}{K_f}\right).
\end{align}
Biot, and others, explored the physical meaning of both $\alpha$ and $M$ in a number of manuscripts \cite{Skempton:1954:a,Biot:1956:c,Biot:1957:a}.
The effective potential, Eq.(\ref{eq:thermodynamic potential}), can also be constructed beginning at microscopic scales to justify the expression through homogenization theory, coarse graining, or mixed micro-macro formulations\cite{Santos:2014:a,Coussy:2004:a,Lopatnikov:2004:a,Nordbotten:2007:a,Bear:2010:a,Apostolakis:2013:a}.

We now use the effective stress and pressure to write the coupled equations of motion for the effective medium and relative fluid flow components in the absence of external stresses.
The equations of motion can be derived via the principle of least action\cite{Lopatnikov:2004:a,Santos:2014:a,Cheng:2016:a}, however that would take us too far afield, so we introduce them qualitatively.
Momentum conservation for the effective medium reads
\begin{equation}\label{eq:bulk EoM}
    (1-\phi)\rho_s\ddot{\bs{u}}_s+\phi\rho_f\ddot{\bs{u}}_f=\nabla\Cdot\bs{\sigma}+(1-\phi)\rho_s\bs{g}+\phi\rho_f\bs{g},
\end{equation}
where the LHS is the total change in inertia of both solid and fluid.
If we define the effective density as $\rho=(1-\phi)\rho_s+\phi\rho_f$, the LHS can be written in a more illuminating form as $\rho\ddot{\bs{u}}_s+\rho_f\phi(\ddot{\bs{u}}_f-\ddot{\bs{u}}_s)$ - the first term is the effective inertia following the acceleration of the solid component, while the second term is the relative acceleration of the fluid with respect to the solid.
The RHS is simple to interpret - the divergence of the bulk stress tensor describes the internal forces, while the external forces consist of a gravitational field.

Similarly, momentum conservation for the fluid reads
\begin{equation}\label{eq:fluid EoM}
    \phi \rho_f \ddot{\bs{u}}_f\!+\!\rho_a(\ddot{\bs{u}}_f\!-\!\ddot{\bs{u}}_s)=-\phi\nabla P \!-\! \phi^2\frac{\eta}{k}(\dot{\bs{u}}_f\!-\!\dot{\bs{u}}_s)\!+\!\phi\rho_f\bs{g}. 
\end{equation}
The LHS here is the change in fluid inertia; however a mass density, $\rho_a$, is added for the fluid moving relative to the solid.
This addition can be derived from a micro-macro approach, and captures the effect of the fluid being slowed down through interactions with the structure of the pores\cite{Bear:1990:a,Bear:2010:a,Cheng:2016:a}.
It is typically written as $\rho_a=(a-1)\phi\rho_f$, with $a$ for packed spheres being related to the porosity as $a=\frac{1+\phi}{2\phi}$ - for glass beads at low frequencies it has been measured as $a\approx1.66\pm0.13$, or $\phi\approx 0.43\pm0.05$, which coincides nicely with the porosity of loosely packed spheres, $0.41\pm0.04$ \cite{Berryman:1980:a,Bonnet:1985:a}.
As in Eq.(\ref{eq:bulk EoM}), the LHS of Eq.(\ref{eq:fluid EoM}) is rewritten using the relative acceleration of the fluid to the solid as $\phi\rho_f\ddot{\bs{u}}_s+(\phi\rho_f+\rho_a)(\ddot{\bs{u}}_f-\ddot{\bs{u}}_s)$.
The RHS has the pressure gradient driving the fluid acceleration, but also includes a viscous dissipation term coming from the relative motion of the fluid with respect to the solid, serving to synchronize the former to the latter.
The porosity factor for the former is due to the fact that the pore pressure is acting on the effective area seen by the fluid.
The $\phi^2$ factor on the latter can be broken down as follows - one factor comes from the dynamic viscosity, which is proportional to the fluid mass, while the second comes from the hydraulic permeability, which is proportional to the effective area seen by the fluid mass.

Note the inverse dependence on hydraulic permeability - the more easily the solid permits fluid flow the less dissipation occurs.
Using the definition of the filtration velocity,  Eq.(\ref{eq:fluid EoM}), can be written as a generalization of Darcy's Law, Eq.(\ref{eq:discharge}),
\begin{equation}\label{eq:Generalized Darcy's Law}
    \bs{q} = -\frac{k}{\eta}\left(\nabla P+\rho_f\ddot{\bs{u}}_s+\frac{\rho_a+\phi\rho_f}{\phi^2}\dot{\bs{q}}-\rho_f\bs{g}\right),
\end{equation}
which now includes the inertial forces coming from the motion of the solid skeleton, as well as the fluid's interactions with the pore space.
Accordingly, we drop the subscript on the solid displacement, $\bs{u}=\bs{u}_s$.

The system of equations (\ref{eq:continuity equation}),(\ref{eq:bulk EoM}),(\ref{eq:fluid EoM}) together with the constitutive relations (\ref{eq:constitutive relation 1}),(\ref{eq:constitutive relation 2}), are a more suitable continuum model of tissues containing a mobile interstitial fluid component than standard viscoelastic continua.
Since MRE subjects tissue to oscillatory driving conditions at constant frequency, the system can be simplified by Fourier transforming into frequency space - linearity assures us that distinct temporal modes evolve independently.
The driving frequency, $\omega$, and all of its harmonics, $\omega_n=n\omega\ \ \forall n\in\mathds{Z}^+$, contribute to the spatial signal.
For each of the d.o.f./hyperparameters $x\in(\bs{u},\zeta,\bs{q},\bs{\sigma},P;\bs{g},\gamma)$, the Fourier decomposition and its inverse read
\begin{align*}
    x(\bs{r},t) &= \sum_{n\in\mathds{Z}} \  \tilde{x}_n(\bs{r})e^{-i\omega_n t}\\
    \tilde x_n(\bs{r})&=\frac{1}{T}\int_0^{T}\!\!\! dt\ x(\bs{r},t)e^{i\omega_n t}.
\end{align*}
The decomposition acts formally by replacing $x\rightarrow\tilde x_n$ and $\partial_t\rightarrow-i\omega_n$.
Decomposing each d.o.f. in the equations of motion, we find the mechanical response at each frequency $\omega_n$ is described by 
\begin{gather}
    i\omega_n\tilde\zeta_n=\nabla\Cdot\ft{q}_n-\tilde{\gamma}_n\\
    -\omega_n^2\rho \ft{u}_n-i\omega_n \rho_f \ft{q}_n=\nabla\Cdot\ft{\sigma}_n+\rho\ft{g}_n\\
    -\omega_n^2\rho_f\ft{u}_n-i\omega_n\rho_f\beta_n^{-1}\ft{q}_n=-\nabla\Cdot\tilde P_n+\rho_f\ft{g}_n\\
    \label{eq:beta}
    \text{ where}\hspace{10pt}\beta_n^{-1} = \frac{\rho_a+\phi\rho_f}{\rho_f\phi^2}+i\frac{\eta}{\omega_n\rho_f k}.\hfilneg
\end{gather}
together with the transformed constitutive relations.
Here, we have introduced the poroelastic $\beta$ factor, a key ingredient in differentiating poroelastic and viscoelastic dynamics.
In a static and constant gravitational field we of course have that $\ft{g}_n=-g\delta_{0n}\hat{\bs{z}}$.

We want to connect the MR images of tissue to this model, so we assume that the MRI phase contrast displacement measurements acquired by MRE trace the motion of the elastic skeleton.
Since fluid gives a strong MR signal, this assumption relies on there being significant amounts of fluid in the disconnected pore space, acting as part of the bulk.
Since fluid pressure within tissue is critical for homeostasis, we should reduce our system of equations further by eliminating all variables except the pair $(\ft{u},\tilde{P})$. 
For analytical simplicity, we choose an isotropic elastic stress tensor parameterized by the standard Lam\'e parameters $\mu$ (shear modulus) and $\lambda$, 
\begin{equation}\label{eq:Hookean material}
    \bs{E}_0=\mu \tbasis{ij}{ij}+\mu \tbasis{ij}{ji}+\lambda\mathds{1}\!\otimes\!\mathds{1}.
\end{equation}
After some lengthy algebra, one finds a vector and scalar family of equations for each harmonic,
\begin{widetext}
\begin{gather}\label{eq:poroelastic eqation 1}
    -\omega_n^2(\rho\!-\!\beta_n\rho_f)\ft{u}_n\!=\!\nabla\Cdot\left[\mu\nabla\!\otimes\!\ft{u}_n\!+\!\mu(\nabla\!\otimes\!\ft{u}_n)^T\right]\!+\!\nabla\left(\lambda\nabla\Cdot\ft{u}_n\!-\!\alpha \tilde P_n\right)\!+\!\beta_n\nabla \tilde P_n\!+\!(\rho\!-\!\beta_n\rho_f)\ft{g}_n\\
    \label{eq:poroelastic equation 2}
    \alpha\nabla\Cdot\ft{u}_n\!-\!\nabla\Cdot[\beta_n\ft{u}_n]=\!-\frac{1}{\rho_f\omega_n^2}\nabla\Cdot\left[\beta_n\nabla \tilde P_n\right]\!-\!\frac{1}{M}\tilde{P_n}\!+\!\frac{1}{\omega_n^2}\nabla\Cdot[\beta_n\ft{g}_n]\!+\!\frac{i}{\omega_n}\tilde\gamma_n.
\end{gather}
\end{widetext}
Up to this point we have not imposed assumptions on the spatial distribution of the elastic moduli in the constitutive relations, or the hydraulic permeability. 
We do so now by assuming that the underlying solid and fluid are nearly incompressible relative to the bulk, namely that $K/K'_s\ll1$, $K/K''_s\ll1$, and $K/K_f\ll1$ \cite{Perrinez:2008:a}.
Plugging the first of these in Eq.(\ref{eq:Biot coefficient}) shows that in this regime $\alpha\rightarrow 1$, while using all three in Eq.(\ref{eq:Biot modulus}) shows that $M^{-1}\rightarrow0$.
In this physical regime, the above reduce to \cite{Detournay:1993:a}
\begin{widetext}
\begin{gather}\label{eq:poro1}
    -\omega_n^2(\rho\!-\!\beta_n\rho_f)\ft{u}_n=\nabla\Cdot\left[\mu\nabla\!\otimes\!\ft{u}_n\!+\!\mu(\nabla\!\otimes\!\ft{u}_n)^T\right]\!+\!\nabla\left(\lambda\nabla\Cdot\ft{u}_n\right)\!-\!(1\!-\!\beta_n)\nabla \tilde P_n\!+\!(\rho\!-\!\beta_n\rho_f)\ft{g}_n\\\label{eq:poro2}
    i\omega_n\rho_f\tilde\gamma_n=\nabla\Cdot\left[(1\!-\!\beta_n)\rho_f\omega_n^2\ft{u}_n\!+\!\beta_n\nabla\tilde P_n\!-\!\beta_n\rho_f\ft{g}_n\right].
\end{gather}
\end{widetext}
The vector equation shares some resemblance to viscoelasticity - the main differences being the presence of the complex $\beta_n$ and displacement driving pressure gradient.
Both Eq.(\ref{eq:bulk EoM}) and Eq.(\ref{eq:fluid EoM}) can be rewritten using the filtration velocity, getting rid of all mentions of the fluid displacement. 
The scalar equation is a divergence condition, implying that the given combination of solid displacement and pressure gradient can only contribute a net curl to the system unless a fluid source is present.
Solving this system with given boundary conditions is referred to as the {\it forward problem}.\\

\subsubsection*{Numerical Considerations I}
Analytical solutions of the equations of motion exist for highly idealized geometries\cite{Coussy:2004:a,Santos:2014:a,Cheng:2016:a}; in the context of classical clinical MRE such idealizations do not exist, hence computational methods are required to find approximate numerical solutions\cite{Lynch:2004:a}.
We use a finite element forward solver, described in detail elsewhere\cite{Perrinez:2010:a,Tan:2017:a}. 
It exploits multiple meshes - displacement fields can be supported on either tetrahedral or hexahedral element meshes, while material properties, such as the shear modulus or hydraulic permeability, are supported on separate hexahedral meshes\cite{Mcgarry:2012:a}.
Supporting each property on its own mesh provides versatility.
It decouples the resolution of the unknown properties being inferred from the mesh resolution of the ($\bs{u},P$) solutions generated by the forward problem, which balances discretization error and computational load without impacting the spatial scale of parameter inference.
As we move forward into work on phantoms and human subjects, this versatility will allow for mitigating the effects of noise through coarser property meshes.\\

\subsection*{Inference}

In MRE we observe the displacement field of the tissue being examined with little or no knowledge of the underlying mechanical properties of the tissue.
This {\it inverse problem} of using the observation to infer the properties, is ideal for the application of a powerful epistemological tool known as Bayesian inference.

The Bayesian interpretation of probability brings with it many tools, some of which have started to be used by the MRE community\cite{Franck:2016:a, Jiang:2020:a}. 
The fundamental idea behind it is simple and intuitive - observation carries with it information, and information leads to a reduction in uncertainty\cite{Cox:1963:a,Jaynes:2003:a,Terenin:2015:a,Sowinski:2016:a,Sowinski:2018:a}.
Beliefs evolve in order to reduce uncertainty, thereby maximizing the fitness of a cognizant being within an environment that is sufficiently complex to be indistinguishable from one behaving pseudorandomly.
Note that, in a Bayesian setting, belief/plausibility are synonymous with probability, and in what follows the terms will be used interchangeably\cite{Van:2003:a,Caticha:2008:a}.

Applying these ideas to MRE, let us fix a particular mechanical model of tissue so that the forward problem is specified.
The space of all possible material property fields is denoted $\mathcal M=\{\bs{\theta}\}$.
An epsitemic agent has a preexisting, or {\it prior}, belief distribution over $\mathcal M$, namely $p(\bs{\theta})$\cite{Gelman:2017:a}.
If they know absolutely nothing about what to expect then the prior is uniform over all possible property fields - a uniform distribution is maximally uncertain\cite{Caticha:2004:a}.
Of course, if they have information that may reduce uncertainty in some way, the prior could be peaked over a region of $\mathcal M$ that is deemed more likely to occur.
The space of all possible observations of the tissue displacement field is denoted $\mathcal D = \{\bs{u}\}$.
Finally, the {\it posterior} over $\mathcal M$ is the informed belief distribution that the epistemic agent has after observing the data, denoted $p(\bs{\theta}|\bs{u})$. 
The information content of a particular belief is defined uniquely\cite{Shannon:1948:a,Cox:1963:a} in order to satisfy a natural sub-additive constraint on dependent beliefs, $I[\boldsymbol{\theta}]=-\log p(\boldsymbol{\theta})$.
Note that strong beliefs entail little information content, while implausible beliefs the opposite.
Observing a rare event endows one with more information than observing a common event. 
It is for this reason that sometimes information content is referred to as {\it surprise} - we will be content in interpreting information content as a measure of the implausibility of a belief.

The posterior and prior are connected via Bayes' theorem
\begin{align}\label{eq:Bayes Rule}
    p(\bs{\theta}|\bs{u})=\frac{p(\bs{u}|\bs{\theta})}{p(\bs{u})}p(\bs{\theta})\nonumber\\
    \equiv\mathcal{L}(\bs{u},\bs{\theta})p(\bs{\theta})
\end{align}
where the quantity multiplying the prior is the {\it likelihood} of the observation given a model.
Note that observations that are more likely in the context of a set of material properties, $\bs{\theta}$, have $\mathcal L(\bs{u},\bs{\theta})>1$ and enhance the posterior over the prior, whereas unlikely observations, $\mathcal L(\bs{u},\bs{\theta})<1$, reduce the belief in that particular $\bs{\theta}$.

Specifying a poroelastic model with parameters $\bs{\theta}=\{\mu(\bs{r}),k(\bs{r})\}$, along with the hyperparameters $\{\phi,\rho_s,\rho_f,\rho_a, \eta,\lambda\}$ and boundary conditions, one solves the equations of motion Eq.(\ref{eq:poro1}),(\ref{eq:poro2}) for $\tilde{\bs{u}}_n(\bs{r})$.
As will be explained shortly, we need only do this for the fundamental frequency, so we define the {\it model displacement} as $\bs{u}^M(\bs{r},\bs{\theta})=\tilde{\bs{u}}_1(\bs{r})$.
We have left the model as an argument here to remind us that the solution for the fundamental mode depends on the parameters that have been chosen.

MRE data acquisition images steady state vibration fields using phase sensitive MRI sequences \cite{Muthupillai:1995a, Manduca:2001:a, Sinkus:2005:b, Sack:2008:b, Johnson:2013:a, Johnson:2014:a}.  
Displacements are measured at a number of times across the harmonic motion cycle, which are processed into a sequence of displacements at each voxel and the fundamental mode of the time series is computed to give $\bs{u}^D(\bs{r})e^{i\omega t}$ for each voxel \cite{Wang:2008:a, Wang:2008:b}.
Vibrations are often in the range of $25-100$Hz and supplied using an external actuator, or alternatively, our intrinsic actuation MRE uses retrospective gating to measure cardiac induced motions at $\sim 1$Hz without the need for actuation hardware \cite{Weaver:2012:a}. 

With $\bs{u}^D$ and $\bs{u}^M$ in hand, we are much closer to understanding the likelihood.
The latter are encoding the material property degrees of freedom, while the former provide constraints.
Both are discretized so one must always make sure to interpolate to the same voxel domain, taking into account that there are enough constraints to fully fix the degrees of freedom.
Once we have fixed the voxel domain, we can treat the scalar difference between the two as a random variable $\Delta u_i(\bs{\theta})=|\bs{u}^D(\bs{r}_i)-\bs{u}^M(\bs{r}_i,\bs{\theta})|$.
The differences are independent across voxels, though this assumption can be relaxed when the noise structure of the data collection can be properly described.
The $\Delta u_i$ are at the voxel level, so each one is actually the mean of all the degrees of freedom at the subvoxel level, of which there are a great many.
Invoking the Central Limit Theorem allows us to claim that the distribution of $\Delta u_i$ is Gaussian for each voxel such that
\begin{align}
    \mathcal L(\bs{u},\bs{\theta})&\propto p(\bs{u}|\bs{\theta})\nonumber\\
    &=\prod_ip(\Delta u_i)\nonumber\\
    &\propto \exp{\left(-\frac{1}{2}\sum_i\frac{\Delta u_i^2}{\sigma_i^2}\right)},
\end{align}
with the $\sigma_i$ quantifying the root mean square fluctuations in the subvoxel degrees of freedom of the $i^{th}$ voxel.
Since no estimate is available for these values, they are set to unity.
Using Bayes' Rule, Eq.(\ref{eq:Bayes Rule}), 
\begin{align}\label{eq:cost function}
    I[\bs{\theta}|\ft{u}]&=-\log\mathcal L(\ft{u}|\bs{\theta})+I[\bs{\theta}]\nonumber\\
    &\simeq\frac{1}{2}\sum_i|\bs{u}^D(\bs{r}_i)\!-\!\bs{u}^M(\bs{r}_i,\bs{\theta})|^2+I[\bs{\theta}],
\end{align}
where the $\simeq$ in the second line means equality modulo a constant independent of $\bs{\theta}$, leads to  Eq.(\ref{eq:cost function}) as a {\it cost function}; the first term on the RHS is an {\it error function}, while the second term, is a {\it regularizer}.
Finding the set $\bs{\theta}$ that minimizes cost is equivalent to having found the most believable set of parameters - an epistemological interpretation grounded in the tools of information theory.\\

\subsubsection*{Numerical Considerations II}
There are, of course, multiple ways to attack this particular question, but they all have in common the theme of sliding down the cost surface.
If the cost surface is convex, such a slide will eventually get us to the most plausible set of parameters.
The idea can be broken down as follows:
Let's say that we are at some arbitrary model $\bs{\theta}\in\mathcal M$.
Our goal will be to move to a new model $\bs{\theta}+\delta\bs{\theta}$ so that the information content at this new model is: $I[\bs{\theta}+\delta \bs{\theta}|\bs{u}]\le I[\bs{\theta}|\bs{u}]$.
We are content to slide either down or horizontally, but never back up the error surface.
By an appropriate sequence of choices for the $\delta\bs{\theta}$'s one has an iterative procedure for finding the minimum.
Different methods rely on expanding the above information content inequality to different orders in $\delta\bs{\theta}$, and then making appropriate approximations with the resulting gradients.

Our MRE platform has several of these methods implemented - descent direction can be computed via Gauss-Newton, Conjugate Gradient, or Quasi-Newton updates\cite{Tan:2017:a}.
Travel distance is modulated by a binary line search once the descent direction is found.
Computation of gradients depends not only on the log likelihood, but also on prior information content - the regularizer.
Multiple regularization terms have been implemented to take into account prior knowledge on material property limits, their spatial distribution, and smoothness\cite{Perrinez:2010:a,McGarry:2013:a,Johnson:2013:b}.

To manage the computational resources required for multiple solutions of the 3D finite element forward problem required for inference, an efficient zoning strategy has been developed\cite{Evenstar:1999:a,Evenstar:2001:a} that breaks up the domain of interest into smaller overlapping zones.
Measured displacements on the boundary voxels of zones provide Dirichlet boundary conditions, leading to a similar inference problem on a smaller scale.
Computational complexity in our MRE-suite for the finite element solution of Eq.(\ref{eq:poro1}) and (\ref{eq:poro2}) on a cubic zone of side length $L$ scales as $\mathcal O(L^{4.6})$, which is less than the $\mathcal O(L^6)$ complexity of LU factorization due to the sparsity resulting from using local operators\cite{McGarry:2008:a}.
Each zone is treated in parallel, computing as many gradient descent iterations as desired.
Property solutions from each zone are collected and stitched together to complete a global iteration\cite{Mcgarry:2012:a}.
Each global iteration begins with a reseeding of the zone structure to limit propagation of bias or boundary noise inherent in a particular zone throughout the inference.

\begin{figure}[p]  
    \centering
    \includegraphics[width=4.00in]{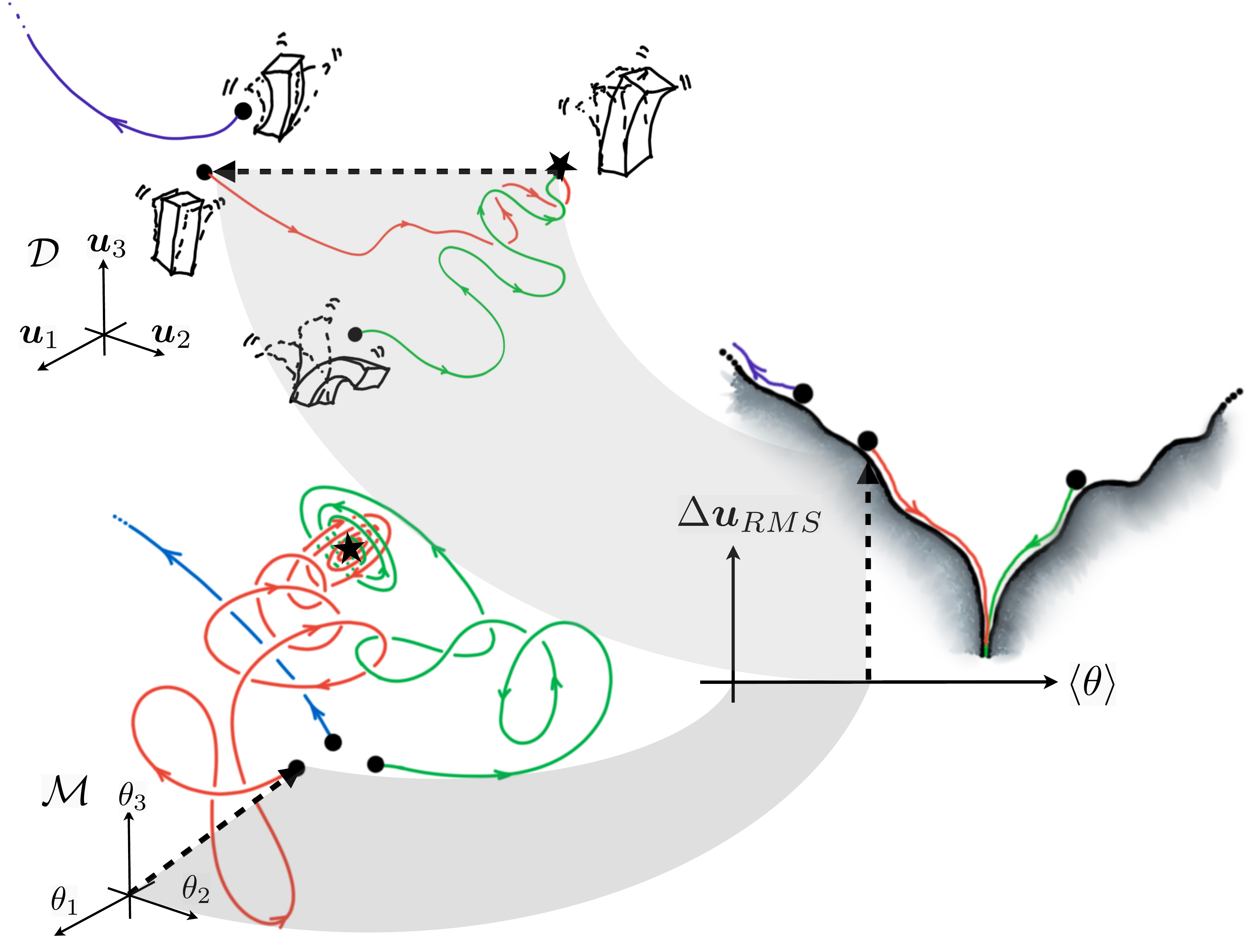}
    \caption{
        A geometric view of the cost surface's lower dimensional representation.
        The trajectories in $\mathcal D$ and $\mathcal M$ (red, green, and blue) are meant to represent how difficult it is to visualize the paths in these high dimensional spaces.
        In MRE, the dimension of both is proportional to the number of voxels.
        To visualize the process of inference, we map our trajectories to a lower dimensional space made of the negative log posterior (root mean square of the displacement errors), and the mean property value. 
        The RMS is proportional to the $L_2$ distance between the data displacements, $\bs{u}^D$, and model displacements, $\bs{u}^M$.
        The mean property value is proportional to the $L_1$ distance from the "no property" origin to the model point, $\bs{\theta}$, .
        During single property inference the resulting lower dimensional representation is $1$-dimensional, while for two property inference it is $2$-dimensional.
        Note how the red and green trajectories eventually leads to the minimal RMS.
        The blue trajectory begins at a high enough error (near vanishing posterior), that our epistemic agent is incapable of finding the correct direction to move in $\mathcal M$. 
        The dimensionality of $\mathcal M$ is so large that nearly every direction leads to an increase in RMS displacement error, and the lower dimensional representation interprets this as rolling up the cost surface, leading to divergence.
        }
        \label{fig:Inference}
\end{figure}

Convergence of any combination of the above methods is tricky to gauge due to the immense dimensionality of $\mathcal M$.
Furthermore, the use of zoning means each global iteration is performing inference on a slightly different cost surface. 
To understand convergence we have therefore opted towards a coarse grained approach, projecting the inferred d.o.f. to a lower dimensional subspace consisting of the mean value(s) of the inferred property descriptors. 
Inference is judged by cost function descent on this subspace.
A geometric interpretation of the method is given in Fig.[\ref{fig:Inference}].

\section{Simulation}\label{sec:3}

\begin{table*}
    \centering
    \begin{tabular}{cccrl}\toprule
        Material Property& Symbol & Value & Units &(Generalized) \\\hline
        Bulk Density    &$\rho$      &$1.0\!\times\!10^3$  &\small{kg/m}$^3$ &$ML^{\!-\!3}$\\
        Fluid Density   &$\rho_f$    &$1.0\!\times\!10^3$  &\small{kg/m}$^3$&$ML^{\!-\!3}$\\   
        Apparent Density    &$\rho_a$    &$1.5\!\times\!10^2$  &\small{kg/m}$^3$&$ML^{\!-\!3}$\\
        Dynamic Viscosity   &$\eta$      &$1.0\!\times\! 10^{-3}$  &Pa$\cdot$s&$ML^{\!-\!1}T^{\!-\!1}$\\
        Porosity    &$\phi$      &$4.0\!\times\! 10^{-1}$  &-\\
        Lam\'e Parameter&$\lambda$   &$1.0\!\times\! 10^4$   &Pa &$ML^{\!-\!1}T^{\!-\!2}$\\\hline
        Shear Modulus   &$\mu_0$     &$3.0\!\times\!10^3$  &Pa&$ML^{\!-\!1}T^{\!-\!2}$\\
        Hydraulic Permeability  &$k_0$       &$1.01\!\times\!10^2$   &D  &$L^2$\\
        \hline
    \end{tabular}
    \caption{The first six material properties were used in both MRE{\it f} and MRE{\it i}. 
    The last two are background values used only in MRE{\it f}. 
    Inclusions have multipliers relative to background values as discussed in the text. 
    In MRE{\it i}, the last two properties, shear modulus and hydraulic permeability, are inferred not specified.}
    \label{table:parameters}
\end{table*}
We use our MRE computational platform - a finite element code capable of simulating displacement fields in poroelastic materials, and an artificial epistemic agent running Bayesian inference on measured displacement fields to extract poroelastic properties. 
The former situation is referred to as the {\it forward problem}, MRE{\it f}, while the latter the {\it inverse problem}, MRE{\it i}.

The following tests examine MRE{\it i}'s ability to reconstruct spatial maps of both the hydraulic permeability, $k$, and the real shear modulus, $\mu$.
First we examined our code's baseline ability to infer distributions of a single property at a time.
Next we considered the more complex requirement of inferring both material property distributions simultaneously.
Simulated data were generated via MRE{\it f}, followed by an exploration of the parameter space to see how initial conditions affect material property inference and find a range of starting conditions which converge to the known ground truth.\\

\subsection*{Single Property Inference}

\subsubsection*{Forward Problem}
\begin{figure}[p]  
    \centering
    \includegraphics[width=6.50in]{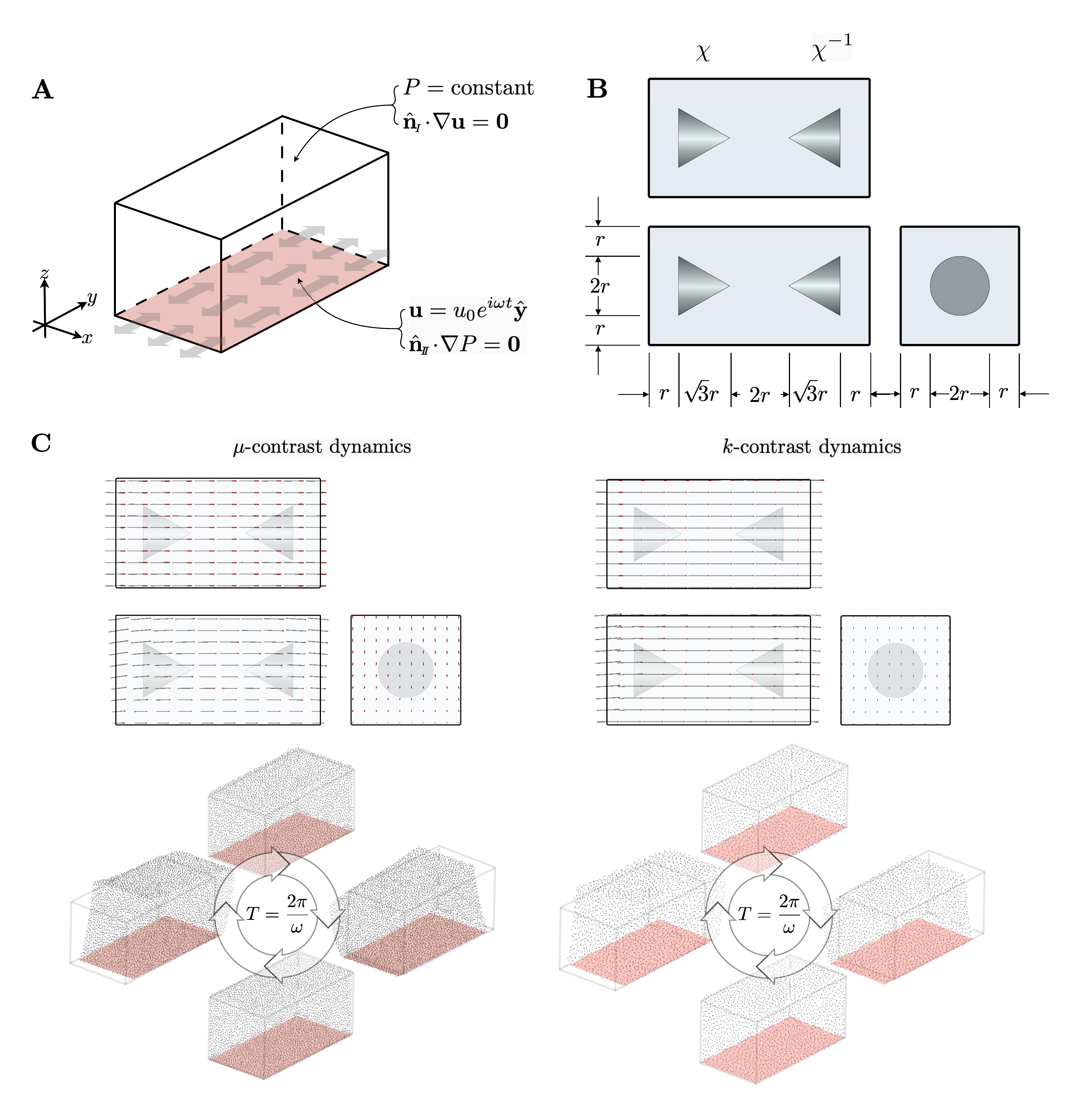}
    \caption{
        \textbf{(A)} 
        Boundary conditions on the top/sides($I$) and bottom($I\!\!I$) surfaces for the $10$Hz simulation. 
        $\hat{\mathbf n}$ are outward pointing normals. 
        Pressure has Dirichlet BCs on $I$ and Neumann BCs on $I\!\!I$. 
        Displacement has the opposite. 
        The bottom plate oscillates in the $y$ direction at frequency $\omega$.
        \textbf{(B)} 
        Orthographic projection showing geometry of the conical inclusions.
        The left inclusion has positive property contrast, and the right one has a negative property contrast.
        \textbf{(C)}
        The dynamics of both scenarios - the left figure has inclusions with shear modulus contrast, while the right figure has inclusions with hydraulic permeability contrast.
        The MRE{\it f} displacement field over one period is shown for both, the red dots representing in-phase displacements at the start of a cycle.
        Below these, the motion of the boundary nodes is displayed at four equally spaced times within the cycle of period $T$.
        }
        \label{fig:1prop_10Hz_setup}
\end{figure}

\begin{figure}[p] 
    \centering
    \includegraphics[width=6.50in]{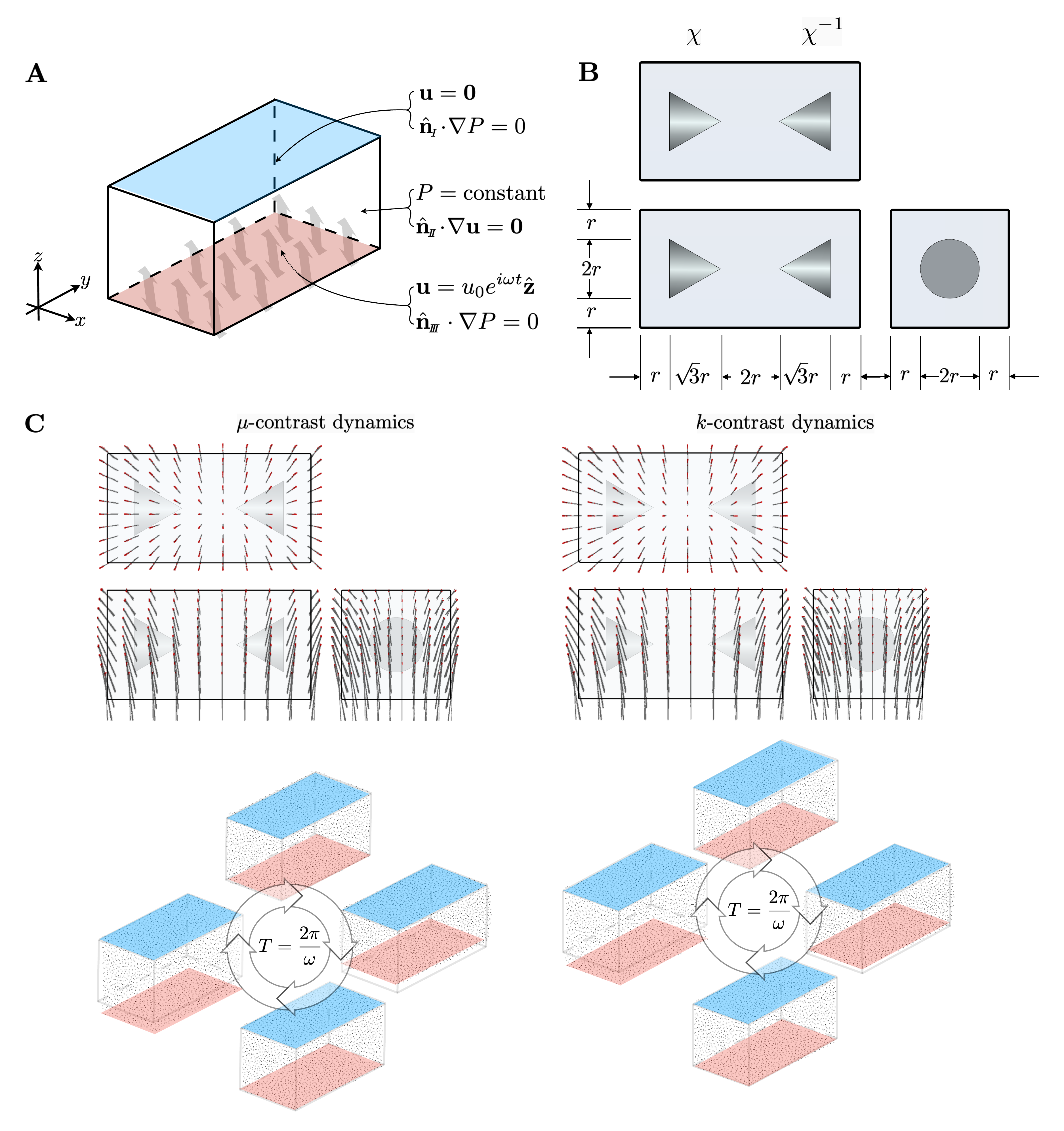}
        \caption{
        \textbf{(A)} 
        Boundary conditions on the top($I$), side($I\!\!I$), and bottom($I\!\!I\!\!I$) surfaces for the $1$Hz simulation. 
        $\hat{\mathbf n}$ are outward pointing normals. 
        Pressure has Neumann BCs on $I$ \& $I\!\!I\!\!I$,   and Dirichlet BCs on $I\!\!I$. 
        Displacement has the opposite. 
        Top plate is held fixed, while the bottom oscillates in the $z$ direction at frequency $\omega$.
        \textbf{(B)} 
        Orthographic projection showing geometry of the conical inclusions.
        The left inclusion has positive property contrast, and the right one has a negative property contrast.
        \textbf{(C)}
        The dynamics of both scenarios - the left figure has inclusions with shear modulus contrast, while the right figure has inclusions with hydraulic permeability contrast.
        MRE{\it f} displacement field over one period is shown for both, the red dots representing in-phase displacements at the start of a cycle.
        Below these, the motion of the boundary nodes is displayed at four equally spaced times within the cycle of period $T$.
        }
    \label{fig:1prop_1Hz_setup}
\end{figure}

Reconstructing the shear modulus or the hydraulic permeability one at a time was tested through configurations depicted in Fig.[\ref{fig:1prop_10Hz_setup}]($10$Hz) and Fig.[\ref{fig:1prop_1Hz_setup}]($1$Hz) - a long rectangular block containing two conical inclusions.
For both frequencies we tested two distinct scenarios.
In the first scenario, inclusions were assigned shear modulus contrast.
Relative to the background shear modulus $\mu_0$, the inclusion on the right has a shear modulus of $\mu = \chi\mu_0$, where $\chi>1$, whereas the one on the left has a shear modulus of $\mu=\chi^{-1}\mu_0$.
The shear modulus {\it contrast} of the right inclusion is positive $(\mu-\mu_0)/\mu_0=\chi-1>0$, while the one on the left is negative, $(\mu-\mu_0)/\mu_0=-(\chi-1)/\chi<0$.
In the second scenario the inclusions were given hydraulic permeability contrast.
We ran simulations with $\chi$ ranging from $1.01$ to $10$, with similar results, and here, we report on the case $\chi=2.0$, in which the inclusions have material property contrasts $-\!1/2$ and $1$ - in the range reported for in vivo soft tissue MRE.

Referencing Fig.[\ref{fig:1prop_10Hz_setup}] and Fig.[\ref{fig:1prop_1Hz_setup}], we chose $r=4.25$mm, making the block dimensions $1.7\text{cm}\!\times\!3.2\text{cm}\!\times\!1.7\text{cm}$.
The displacements were supported on a tetrahedral mesh with $10,081$ vertices and $55,593$ elements.
Although this is smaller than typical organs imaged with MRE in practice, results were similar to larger values of $r$, and the computational cost of running a large number of simulations favored smaller geometries for detailed numerical investigations.   
Material properties were supported on a $0.5$mm cubic lattice mesh with $71,672$ vertices and $66,429$ elements.
The conical inclusions each have a relative volume of $V_c/V=\pi(2\sqrt{3}-3)/32\approx4.56\%$.
The displacement mesh intersects the inclusions at $453$ and $461$ vertices for the left and right cones, respectively.

We report results the results for both frequencies,  $f=\omega/2\pi=1$Hz and $10$Hz.
For the shear modulus frequencies between $1\!\!-\!\!100$Hz were tested with similar results.
For hydraulic permeability we evaluated a smaller range of frequencies, finding that the reconstruction worsened at frequencies above $10$Hz.
At $10$Hz, see Fig[\ref{fig:1prop_10Hz_setup}], fixed displacement boundary conditions were imposed on the bottom face of the block to generate oscillations parallel to its long axis, $\boldsymbol{u}=u_0e^{i\omega t}\hat{\boldsymbol{y}}$.
At $1$Hz, see Fig[\ref{fig:1prop_1Hz_setup}], oscillations perpendicular to the bottom face were imposed, $\boldsymbol{u}=u_0e^{i\omega t}\hat{\boldsymbol{z}}$
The displacement amplitude in both cases was chosen to be $0.5$mm, justifying a linear treatment, $u_0/L\approx0.036\ll 1$.

Material property values used in MRE{\it f} are summarized in Table \ref{table:parameters}.
Bulk density was chosen to be similar to brain tissue, while fluid density and dynamic viscosity were selected to fall in the range of cerebrospinal fluid\cite{Barber:1970:a,Cala:1981:a,Diresta:1990:a,Bloomfield:1998:a,Yatsushiro:2018:a}.
The apparent density is a coarse grained effective quantity coming from the interactions between the fluid and the pore space, and we chose a value based on prior studies, noting that a low sensitivity of results to $\rho_a$ was demonstrated \cite{Perrinez:2007:a,Perrinez:2008:a,Perrinez:2010:a,Pattison:2014:a}.
Lam\'e constants have been well studied in MRE, our choices reflecting reasonable values\cite{Johnson:2013:b}.
Porosity was chosen based on mean values for white and grey matter using DTI, as was the background hydraulic permeability\cite{Venton:2017:a}.
It should be noted that these last two properties are not well measured, with the literature quoting values in a wide range \cite{Cheng:2007:a,Chen:2007:a,Stoverud:2009:a,Ray:2019:a}.\\

\subsubsection*{Inverse Problem}

Inference uses a coarser material property mesh in order to ensure enough constraints to solve the equations of motion.
A side effect of the forward and inverse meshes not being identical is that there is no zero error solution. 
Here we have a $2.0$mm cubic lattice mesh, with $1700$ vertices and $1296$ volume elements.

We explore the question of how close to the true mean our initial values have to be in order to ensure convergence by examining ranges spanning $\sim2$ orders of magnitude above and below the true mean.
The background shear modulus is set to $\mu_0=3.0$kPa, and the background hydraulic permeability is $k_0=101.3$D.
Thus, the ground truth mean values for $\chi = 2$  are:
\begin{align*}
    \langle\mu\rangle_{GT}&=\left(1+\frac{V_c}{V}\frac{(\chi-1)^2}{\chi}\right)\mu_0\approx 3068\text{Pa}\\
    \langle k\rangle_{GT}&=\left(1+\frac{V_c}{V}\frac{(\chi-1)^2}{\chi}\right)k_0\approx 103.6\text{D}
\end{align*}
These expressions are, of course, only approximate for a discrete mesh - deviations on the order of $1\%$ were observed for our constructed meshes.
Since these values are so close to the background fields due to the tiny volume occupied by the inclusions, we considered an inversion successful if the measured mean property fell near the region between the background and mean property values.

We initialize inference with a homogeneous material property distribution of either $\langle \mu\rangle$ or $\langle k\rangle$.
Twenty $(20)$ values of shear modulus were chosen ranging an order of magnitude above and below the true mean, $150-30,000$Pa.
Twenty $(20)$ values of hydraulic permeability were chosen ranging nearly two orders of magnitude above and below the true mean, $10^0-10^4$D.

The inversion used zones with $10\%$ overlap. 
Zones were large enough that approximately $500$ non-boundary nodes existed in each zone, with about $60$ nodes in each zone overlapping with other zones.
Gaussian smoothing regularization (spatial filtering) was applied, with the smoothing radius shrinking from $5$mm to $0.5$mm over the first $50$ iterations and held constant for the remaining iterations.
A conjugate gradient descent method was used for iterative inference on each zone. 
During the first $20$ global iterations, each zone finds a single CG direction, and the distance moved is computed through a single iteration of a secant linesearch.
The next $30$ iterations use two of each, and the remaining use three. 

Results of inference are summarized for the $10$Hz data in Fig.[\ref{fig:1prop_10Hz_mu}] for $\mu$ and Fig.[\ref{fig:1prop_10Hz_hp}] for $k$.
For the $1$Hz data, the shear modulus results are shown in Fig.[\ref{fig:1prop_1Hz_mu}] and the hydraulic permeability in Fig.[\ref{fig:1prop_1Hz_hp}].\\

\subsection*{Double Property Inference}

\subsubsection*{Forward Problem}

\begin{figure}[p] 
    \centering
    \includegraphics[width=7.00in]{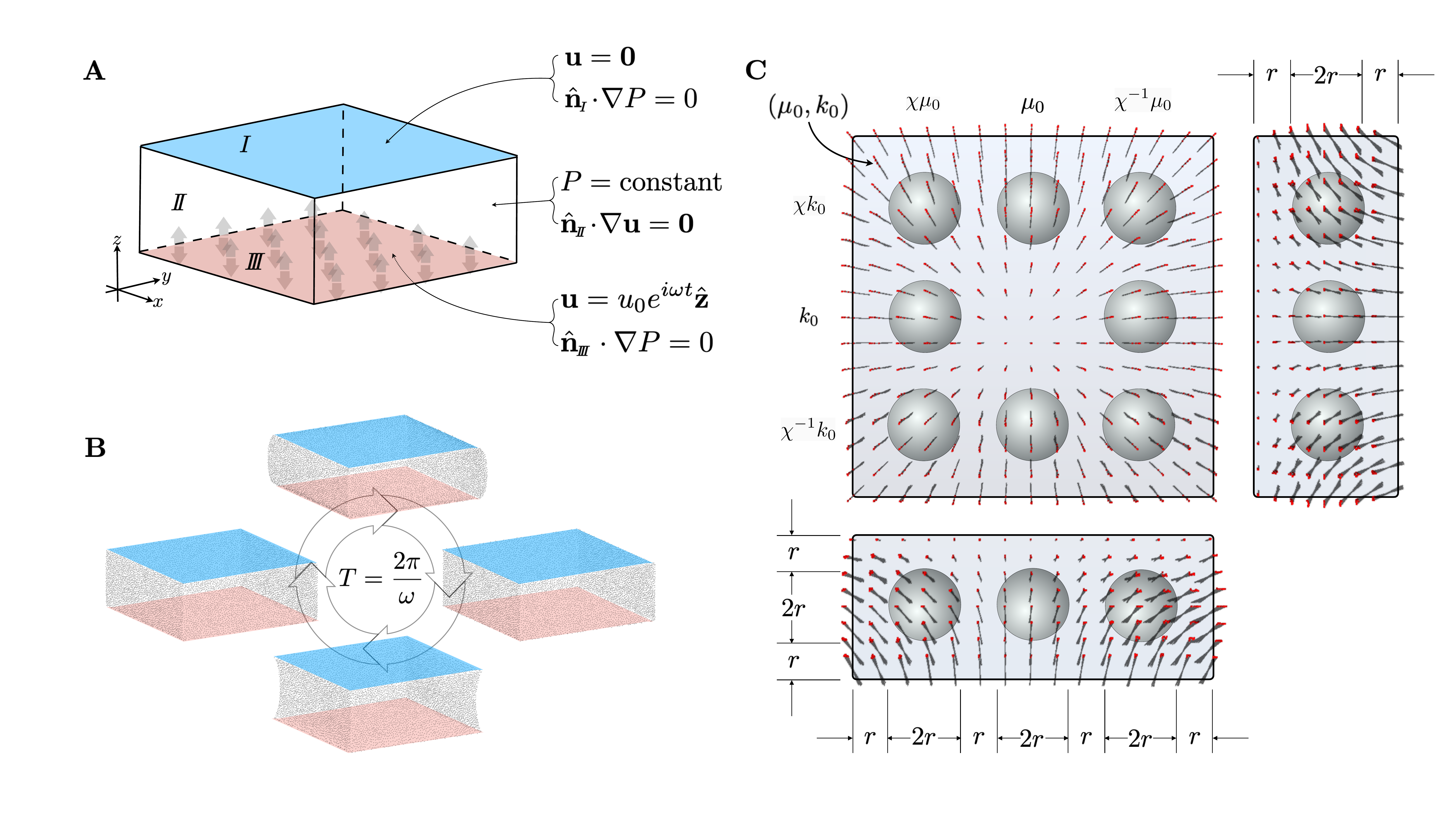}
        \caption{
        \textbf{(A)} 
        Boundary conditions on the top($I$), side($I\!\!I$), and bottom($I\!\!I\!\!I$) surfaces for the two property inference simulation at $1$Hz.. 
        $\hat{\mathbf n}$ are outward pointing normals. 
        Pressure has Neumann BCs on $I$ \& $I\!\!I\!\!I$,   and Dirichlet BCs on $I\!\!I$. 
        Displacement has the opposite. 
        Top plate is held fixed, while the bottom oscillates in the $z$ direction at frequency $\omega$.
        \textbf{(B)}
        The motion of the boundary nodes is displayed at four equally spaced times within the cycle of period $T$.
        One can see the Poisson effect during the apexes of the squeeze/stretch cycles.
        \textbf{(C)} 
        Orthographic projection showing the geometry of the eight spherical inclusions.
        The left inclusions have positive shear modulus contrast, while the right ones has negative contrast.
        The top inclusions have positive hydraulic permeability contrast, while the bottom ones have negative.
        The MRE{\it f} displacement field over one period is superimposed, the red dots representing in-phase displacements at the start of a cycle.
        }
    \label{fig:2prop_setup}
\end{figure}

MRE{\it f} simulation of combined shear modulus and hydraulic permeability contrast involved the configuration depicted in Fig.[\ref{fig:2prop_setup}] - a square block containing eight inclusions.
The inclusions are spherically symmetric, and have material property multipliers of either $1,\chi,$ or $\chi^{-1}$, resulting in either null, positive, or negative material property contrasts, respectively.
Corner inclusions had both properties contrasted, while the center inclusions have contrast in one. 
This setup tests all possible pairs of positive, negative, or null contrasts over all the inclusions.
Multiple values of $\chi$ were tested ranging from $1.1-3.0$, with similar results - we report the case $\chi=2.0$, corresponding to inclusions with property contrasts of either $1$ or $-1/2$.

Referring to Fig.[\ref{fig:2prop_setup}], We chose $r=3.5\text{mm}$, making the block dimensions  $3.5\text{cm}\!\times\!3.5\text{cm}\!\times\!1.4\text{cm}$.
MRPEf used a $1$mm tetrahedral displacement mesh ($23,275$ vertices and $132,084$ volume elements) and a $1.0$mm cubic lattice material property mesh ($23,104$ vertices and $20,535$ volume elements).
Each inclusion took up $\pi/300\approx 1\%$ of the total volume, containing $\sim\! 240$ vertices and $\sim\!215$ elements - deviations from inclusion smoothness occur at $\sim\!0.35$ steradians.

Boundary conditions were imposed to simulate the top face of the block being held fixed, while the bottom face was vibrated in the vertical direction, $\mathbf u=u_0e^{i\omega t}\hat{\mathbf z}$.
The amplitude was chosen to be $u_0=0.5$mm, ensuring that the size of vibrations relative to block geometry is small, $u_0/H =1/28\approx 0.04\ll 1$, so that a linear treatment is justified.
A frequency of $f=\omega/2\pi=1$Hz was applied.
As in the case of single property inference, material properties were chosen to reflect known values associated with brain tissue. 
Our choices are summarized in Table \ref{table:parameters}.\\

\subsubsection*{Inverse Problem}
MRE{\it i} deployed a coarser material property mesh than MRE{\it f} - a $2.5$mm cubic mesh with $1792$ vertices and $1350$ elements. 
Each inclusion sampled $12-13$ material mesh nodes.

We examined the effects of initial conditions on reconstruction fidelity by applying $20$ different homogeneous initial conditions around the mean values, with the shear modulus ranging between $9.13\!\times\! 10^2-5.15\!\times\!10^3$Pa, and hydraulic permeability ranging between $8.9\!\times\!10^{-2}-2.7\!\times\!10^5$D.
Background shear modulus and hydraulic permeability were $\mu_0=3$kPa and $k_0=101.2$D.
Ground truth mean values were
\begin{align*}
    \langle\mu\rangle_{GT}=\left(1+\frac{\pi}{100}\frac{(\chi-1)^2}{\chi}\right)\mu_0\approx 3,047\text{Pa}\\
    \langle k\rangle_{GT}=\left(1+\frac{\pi}{100}\frac{(\chi-1)^2}{\chi}\right)k_0\approx102.9\text{D}
\end{align*}
Once again, these were approximate values on the discrete mesh, but fell within $\sim2\%$ of the observed values.
Inference was considered successful if both mean property values fell within or near the region $[k_0,\left<k\right>_{GT}]\!\times\![\mu_0,\left<\mu\right>_{GT}]$.

The first $10$ global iterations used a single CG direction and a single secant line search for computing total distance traversed in parameter space while the next $30$ applied two of each, and the final $250$ invoked three of each.
The first $50$ iterations used Gaussian smoothing regularization (spatial filtering) on both properties, with the Gaussian standard deviation shrinking linearly from $5$mm to $0.5$mm.
If the change in the cost fell below $10^{-12}$, inference was considered complete, and further iterations ceased.

\begin{figure}[p]
    \centering
    \includegraphics[width=6.69in]{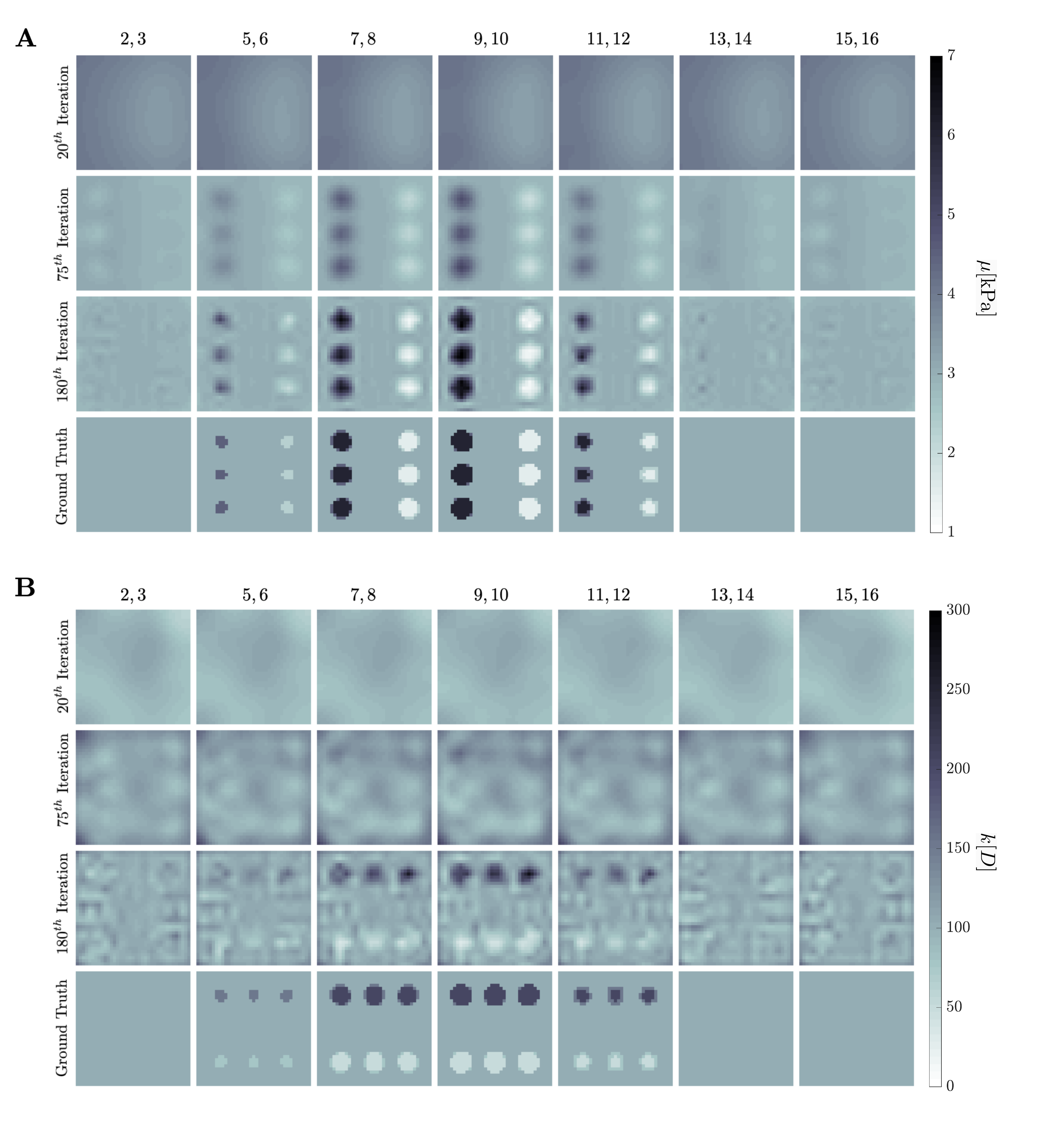}
        \caption{
        Simultaneous inference evolution of both \textbf{(A)} shear modulus, $\mu$, and \textbf{(B)} hydraulic permeability, $k$, in the $1$Hz experiment setup in Fig.\ref{fig:2prop_setup}.
        The former is measured in kilo-Pascal, while the latter in Darcy. 
        Columns are averages of the numbered $(xy)$-planes as in the other simulations.
        Rows are labeled by which iteration they are in MRE{\it i}.
        }
    \label{fig:2prop_recon}
\end{figure}

\begin{figure}[p]
    \centering
    \includegraphics[width=6.69in]{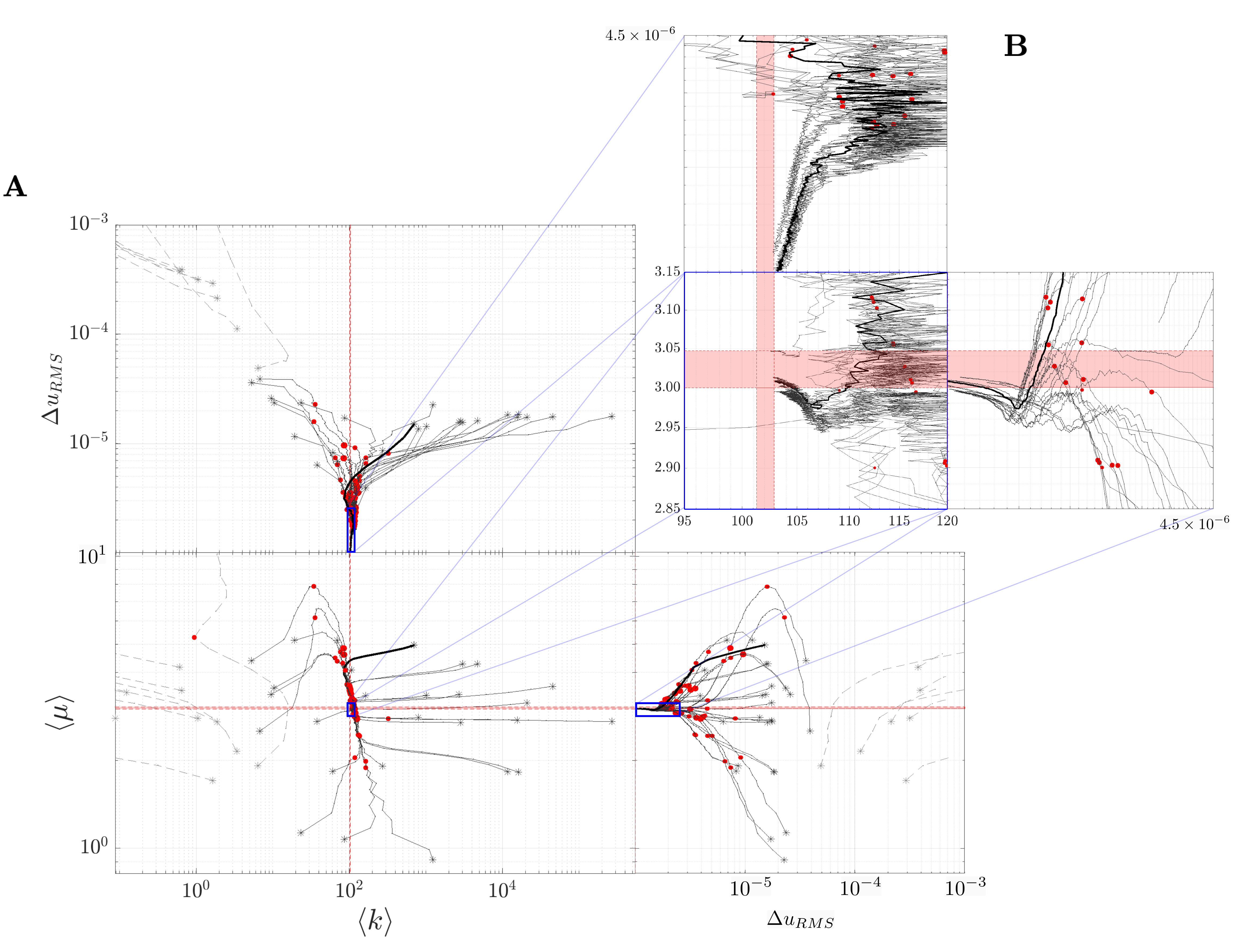}
        \caption{
        Lower dimensional representation of the inference trajectories.
        Note that because two properties are being reconstructed, the lower dimensional representation is $2$-dimensional.
        This plot conveys the trajectories through an orthographic projection.
        (\textbf{A}) shows a large region of $(\langle k\rangle,\langle\mu\rangle)$-space, including starting points of the trajectories represented by asterisks.
        Right and top panels represent root mean square values for converging trajectories.
        Some diverging trajectories are also shown - they are dashed and lighter. 
        The single thick trajectory is the one used for fig.\ref{fig:2prop_recon}.
        Small and large red markers represent points where the descent strategy changed.
        The light red horizontal and vertical regions are the critical strips laying between the background property values and the ground truth mean values.
        Note that, as before, scales are all logarithmic.
        (\textbf{B}) is zoomed into the blue squares located in the figures described above.
        The shaded red areas represent the region between the background property value, $\mu_0$ or $k_0$, and the mean value of the ground truth distributions, $\langle\mu\rangle_{GT}$ or $\langle k\rangle_{GT}$.
        Note that in these plots the scales have changed to be linear.
        Discussion of trends is in the text.
        }
    \label{fig:2prop_evolution}
\end{figure}

Evolution of image reconstruction is summarized in Fig.[\ref{fig:2prop_recon}] and the lower dimensional representation of the inference trajectory is in Fig.[\ref{fig:2prop_evolution}].

\section{Discussion and Conclusion}\label{sec:4}

\begin{figure}[p]
    \centering
    \includegraphics[width=7.00in]{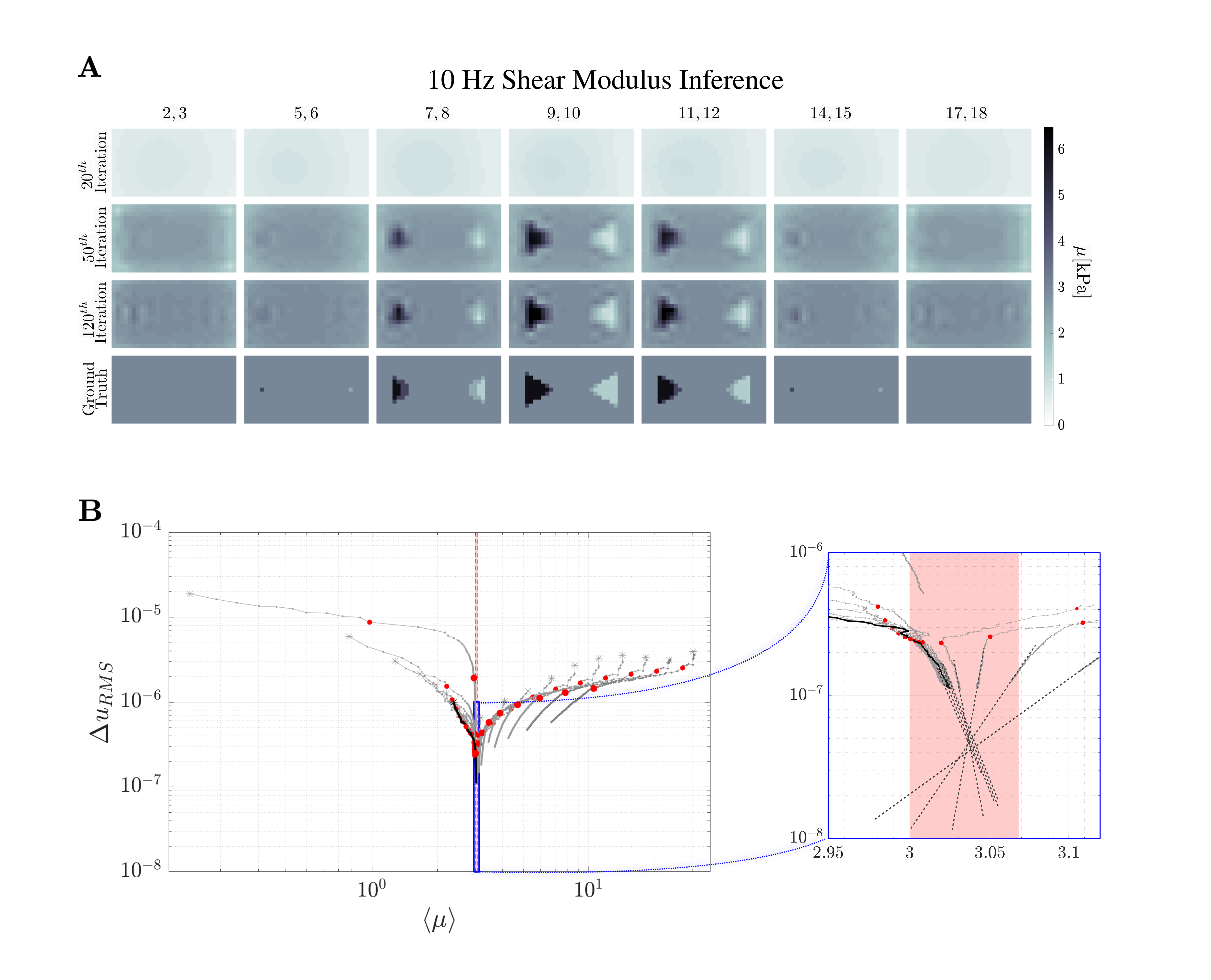}
    \caption{
        Reconstruction evolution of the shear modulus in the 10 Hz simulation.
        (\textbf{A}) shows the reconstructed images of the shear modulus distribution inside the simulated phantom.
        The column images are averages of the numbered $(xy)$-slices (top row), each of which is $2$mm thick.
        The rows are labeled by the iteration number of the inference.
        (\textbf{B}) shows the lower dimensional trajectory of the shear modulus.
        As above, the shear modulus is measured in kilo-Pascals.
        The smaller diagram is a magnified region near the converged values. 
        Initial estimates are noted by an asterisk and changes in descent strategy are indicated by small and large red markers, as explained in the text.
        The light red region represents the strip between the background value of the shear modulus and the ground truth mean value.
        The thicker trajectory is the one used to make the images in the above panel.
        Due to a rather long convergence time, best fit linear extrapolations of the trajectories are plotted to show that the convergence region is (nearly) the same for all converging runs.
    }
    \label{fig:1prop_10Hz_mu}
\end{figure}

\begin{figure}[p]
    \centering
    \includegraphics[width=7.00in]{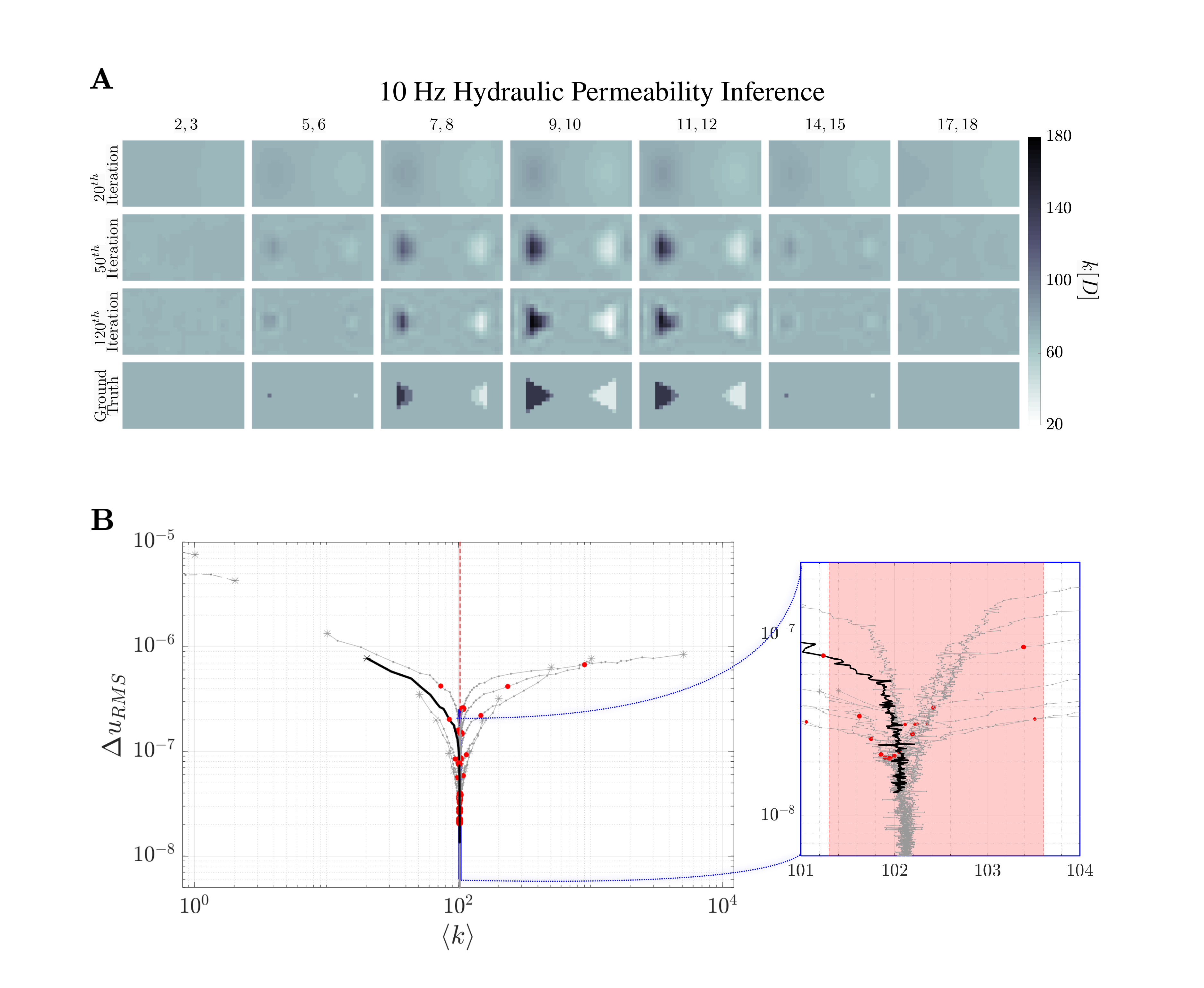}
    \caption{
        Reconstruction evolution of the hydraulic permeability, $k$, at 10 Hz.
        (\textbf{A}) shows the reconstructed images of the hydraulic permeability distribution.
        The column images are the averages of the numbered $(xy)$-planes (top row).
        The rows are labeled by the iteration number of the inference.
        (\textbf{B}) shows the lower dimensional trajectory of the shear modulus.
        As above, the hydraulic permeability is measured in Darcy.
        The smaller diagram is a magnified region near the converged values. 
        Initial estimates are noted by an asterisk and changes in descent strategy are indicated by small and large red markers, as explained in the text.
        The light red region represents the strip between the background value of the shear modulus and the ground truth mean value.
        The thicker trajectory is the one used to make the images in the above panel.
        One can see that the convergence of the permeability occurs well before the final change in descent strategy.
        It appears that proper inference is robust to initial conditions 2 orders of magnitude greater than the ground truth, while only 1 order of magnitude below.
    }
    \label{fig:1prop_10Hz_hp}
\end{figure}

The dynamical evolution of our MRE{\it i} code for single property inference and image reconstruction is displayed in Fig.\ref{fig:1prop_10Hz_mu} and Fig.[\ref{fig:1prop_10Hz_hp}] for $f=10$Hz data.
The top panels of both display reconstructed images at four distinct iterations.
By the $20^{th}$ iteration, both hydraulic permeability and shear modulus are not very distinct, though the former is clearly partitioning the domain into positive and negative contrast regions. 
By the $50^{th}$ iteration, both material properties show distinct localization that is similar to the actual inclusions.
At this stage, spatial filtering has reached sub-voxel width; hence, fuzziness of the hydraulic permeability image reconstruction relative to the sharper shear modulus image reconstruction is solely due to the geometry of the information content surface. 

\begin{figure}[p]
    \centering
    \includegraphics[width=7.00in]{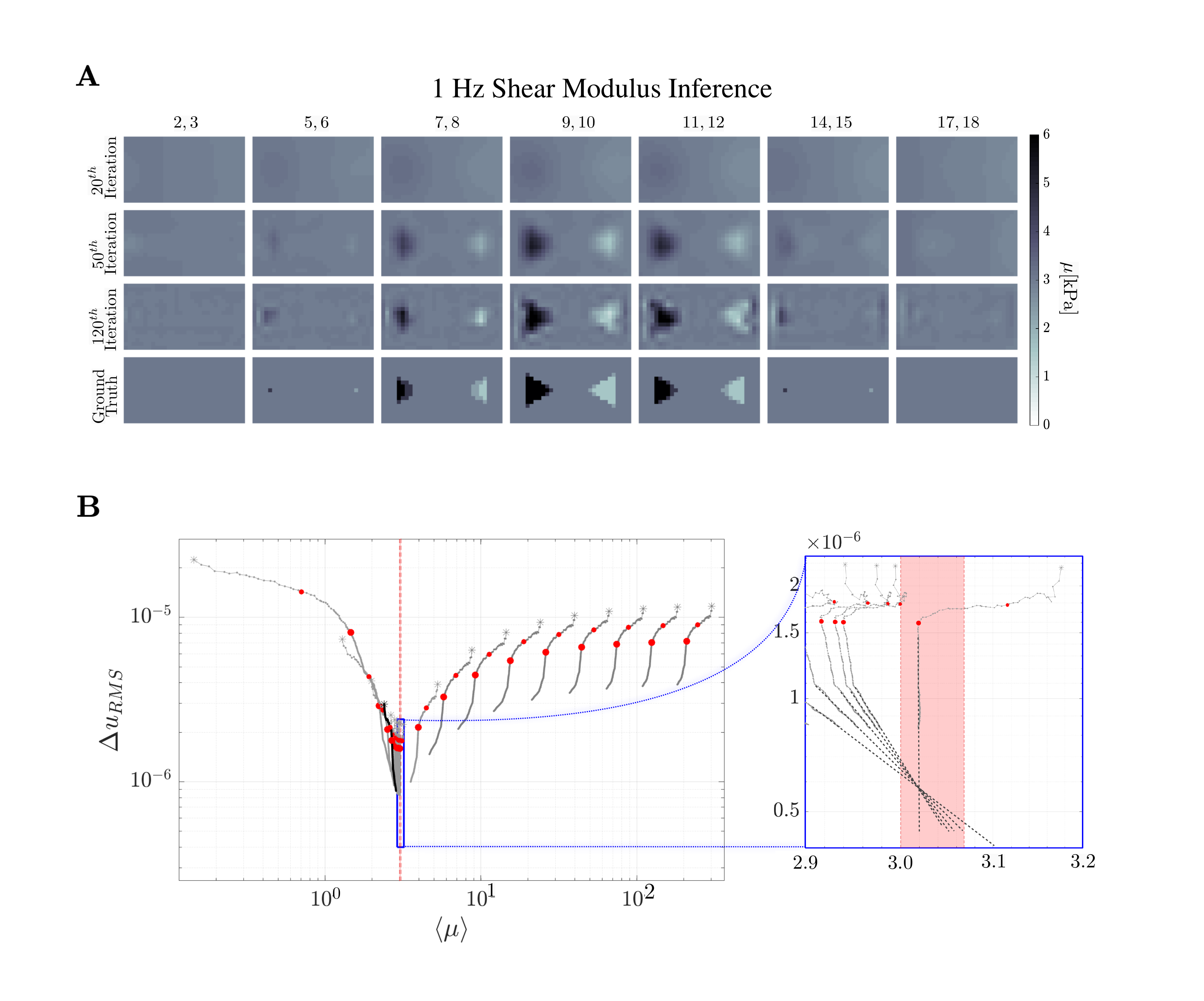}
        \caption{
        Reconstruction evolution of the shear modulus, $\mu$, at 1 Hz.
        (\textbf{A}) shows the reconstructed images of the shear modulus distribution inside the simulated phantom.
        The column images are the averages of the numbered $(xy)$-planes (top row).
        The rows are labeled by the iteration number of the inference.
        (\textbf{B}) shows the lower dimensional trajectory of the shear modulus.
        As above, the shear modulus is measured in kilo-Pascals.
        The smaller diagram is a magnified region near the converged values. 
        Initial estimates are noted by an asterisk and changes in descent strategy are indicated by small and large red markers, as explained in the text.
        The light red region represents the strip between the background value of the shear modulus and the ground truth mean value.
        The thicker trajectory is the one used to make the images in the above panel.
        Once again we see that convergence of shear modulus is much slower than hydraulic permeability, yet images for a wide range of these simulated phantoms all show the correct structure of inclusions.
        Best fit lines are added to the zoomed in image to show that all the trajectories are approaching the same value.
    }
    \label{fig:1prop_1Hz_mu}
\end{figure}

\begin{figure}[p]
    \centering
    \includegraphics[width=7.00in]{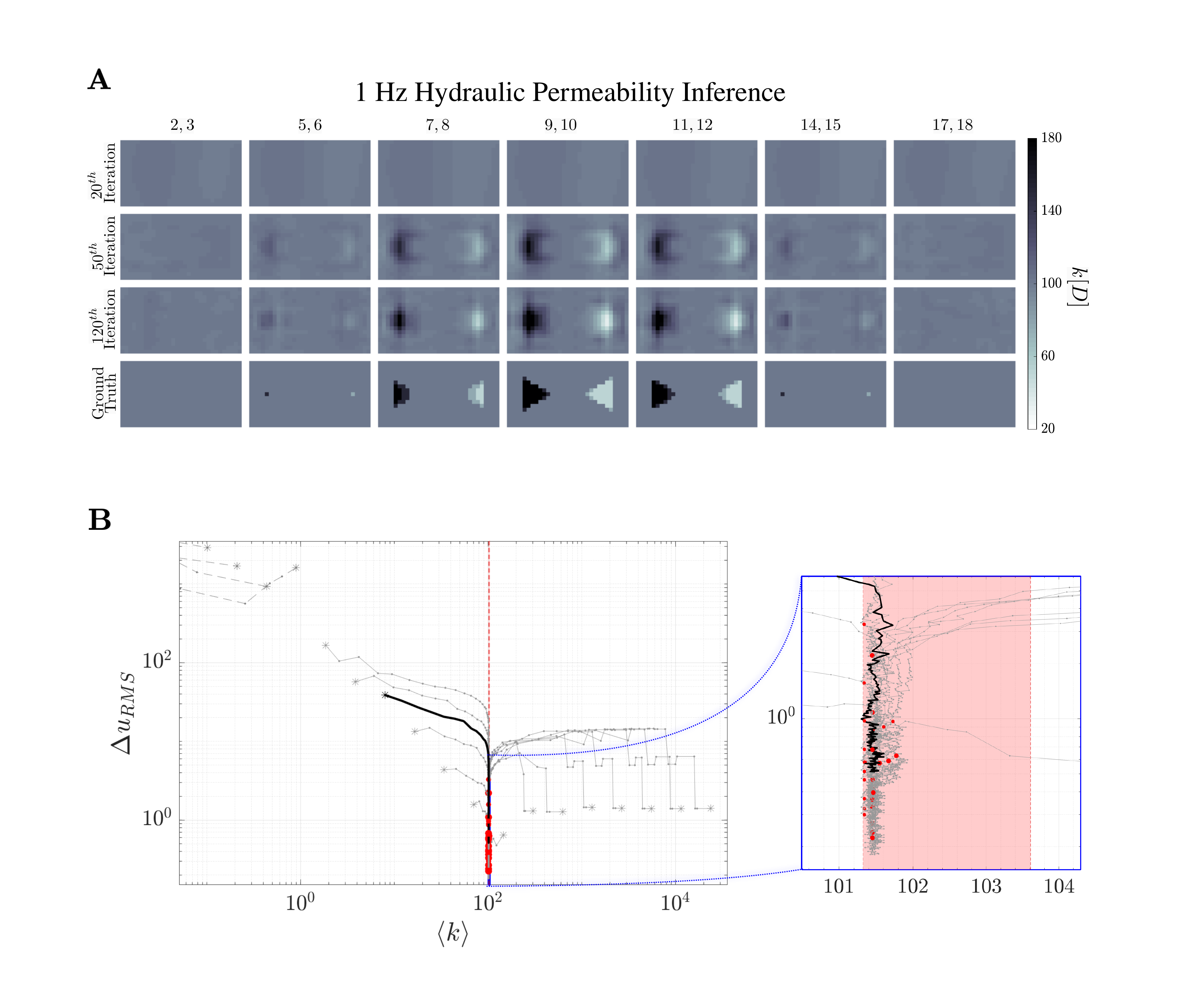}
        \caption{
        Reconstruction evolution of the hydraulic permeability, $k$, at 1 Hz.
        (\textbf{A}) shows the reconstructed images of the hydraulic permeability distribution.
        The column images are the averages of the numbered $(xy)$-planes (top row).
        The rows are labeled by the iteration number of the inference.
        (\textbf{B}) shows the lower dimensional trajectory of the shear modulus.
        As above, the hydraulic permeability is measured in Darcy.
        The smaller diagram is a magnified region near the converged values. 
        Initial estimates are noted by an asterisk and changes in descent strategy are indicated by small and large red markers, as explained in the text.
        The light red region represents the strip between the background value of the shear modulus and the ground truth mean value.
        The thicker trajectory is the one used to make the images in the above panel.
        Once again we witness that hydraulic permeability inference is robust to initial conditions over two orders of magnitude above the ground truth.
        With this simulation we see that the lower bound has gotten better as well, robust out to 2 orders of magnitude below the ground truth.
    }
    \label{fig:1prop_1Hz_hp}
\end{figure}

The $f=1.0$Hz reconstructions are summarized in Fig.[\ref{fig:1prop_1Hz_mu}] and Fig.[\ref{fig:1prop_1Hz_hp}].
Hydraulic permeability starts to settle to the correct values by the $50^{th}$ iteration, though nearly $100$ more iterations are needed for the correct shape of the inclusions to appear in part. This is due to the coarseness of the MRE{\it i} mesh. 
The $\left<k\right>$ value enters the critical strip typically within the first 20 iterations.
We do note that when the initial values of the hydraulic permeability fall two orders of magnitude below ground truth values, MRE{\it i} begins to diverge.
Similar difficulties were not found if initial values overestimated the true hydraulic permeability.
The shear modulus did not show this behavior, with image reconstruction recovering inclusion geometry and relative distribution very well by the $50^{th}$ iteration, although different initial conditions are seen to lead to different final absolute values - a peculiarity noted in prior work\cite{McGarry:2019:a}.
The lower panels of Fig.[\ref{fig:1prop_10Hz_mu},\ref{fig:1prop_10Hz_hp},\ref{fig:1prop_1Hz_mu},\ref{fig:1prop_1Hz_hp}] show the inference trajectories, with initial estimates noted by an asterisk and changes in descent strategy indicated by red markers.

Projecting onto the mean field error surface, hydraulic permeability approaches the correct mean field much earlier than shear modulus. 
Initial values do not affect the late stage reconstruction of the former when they are $\sim2$ order of magnitude too small or up to $\sim3$ orders of magnitude too high.
The CG descent method does appear to have pathological behavior when initial values are more than $\sim2$ order of magnitude too small, resulting in an ascent of the error surface. 
A possible reason for this outcome is that low values of hydraulic permeability induce incompressible elastic behavior known to have stability issues in viscoelastic simulations\cite{Tan:2017:a}.
The geometry of shear modulus reconstruction occurs with high fidelity within very few iterations, but the values converge much slower to the ground truth, an observation reported in prior work\cite{McGarry:2019:a}.
Projected trajectories of the error evolution in both Fig.[\ref{fig:1prop_10Hz_mu}] and Fig.[\ref{fig:1prop_1Hz_mu}], however, converge at or near the critical strip for all initial conditions.

The results of the simultaneous property convergence appear in Fig.[\ref{fig:2prop_recon}] and Fig.[\ref{fig:2prop_evolution}].
The former shows the fidelity of image reconstruction, while the latter inference trajectories as material properties are updated at each global iteration.
Interestingly, the hydraulic permeability enters its critical strip well before the the shear modulus.
Mean value convergence, however, is not a strong indicator of image fidelity.
The shear modulus inclusion geometry is discernible by the $75^{th}$ iteration, whereas nearly twice as many iterations are required before hydraulic permeability inclusion geometry begins to resemble ground truth.
Furthermore, dependency between the two properties can be observed near the edges of the simulation - hydraulic permeability reconstruction has some vertical patterning that resembles the shear inclusions.

Interestingly, we find that the $k$-reconstruction trajectories always approach the critical strip from above. 
Inference tends to overestimate hydraulic permeability with several iterations, and then drives it down.
As in the case of single property inference, $k$ initial conditions have a lower limit of about $\sim1$ order of magnitude below the ground truth in order for MRE{\it i} to converge.
Shear modulus is underestimated in almost all cases within about $50$ iterations, and is then slowly driven upwards to the critical strip, not unlike the inference dynamics for the single property $(\mu)$ case.  

Prior work\cite{Pattison:2014:a,McGarry:2015:a} has performed similar dual property reconstructions.
These results, however, had MRE{\it i} initial conditions very close to ground truth values, utilized the same property mesh for MRE{\it i} as used for MRE{\it f}, and required large amount of regularized smoothing. 
Our work shows that the first two assumptions can be discarded, and the third relaxed significantly, while maintaining high fidelity for image reconstruction.

There seems to be some confusion in the literature between the terms hydraulic conductivity and hydraulic permeability, with many authors using the two interchangeably. 
We focused early in Section \ref{sec:2} on making a clear distinction between the two in order to emphasize the specific contributions to fluid flow coming from the viscosity of the fluid and the geometry of the solid. 
Cerebrospinal fluid has a viscosity ranging between $0.7-1.0$ mPa $\Cdot$ s \cite{Bloomfield:1998:a}, blood between $3.0-67.7$ mPa $\Cdot$ s \cite{Schmidt:1971:a}, ascitic fluid between $0.5-1.5$ mPa $\Cdot$ s \cite{Gokturk:2010:a} - the range covers two orders of magnitude so tissues with similar hydraulic permeabilities can have vastly different hydraulic conductivities.
Hydraulic permeability measurements correlated with pathology can therefore be connected to changes in the {\it elastic matrix} of the tissue (via its effective continuum).

Our motivation for simulated values of porosity and hydraulic permeability rests on a general lack of knowledge by the scientific community of in-vivo values of these parameters.
In vitro experiments tend to find small values for both, which we hypothesize is due to pore collapse in sampled tissue.
It is easy to imagine that once tissue is removed from the body, the pressure change and unjacketing of the sample results in fluid being expelled.
The drop in pore pressure is such that the stresses in the elastic matrix are incapable of supporting the pore space under the influence of gravity, leading to collapse.
Once pores are closed, surface to surface cellular interactions may prevent the process from being reversible.
Even simple models like the Carman-Kozeny\cite{Carman:1937:a,Kozeny:1927:a} relationship between hydraulic permeability and porosity, $k\propto\phi^3$, would predict drastic changes in the former under reasonable changes in the latter.
Furthermore, preliminary work shows that higher values of hydraulic permeability result in higher likelihood, giving us confidence that in vitro measurements of permeability are being underestimated by up to several orders of magnitude.
A full analysis is forthcoming.

The shear modulus, on the other hand, is much better understood, with far less variance in the literature.
In viscoelastic models, white matter typically falls in the range of $1.5-2.5$kPa, depending on frequency, while gray matter is between $0.75-1.5$kPa\cite{Budday:2015:a,McGarry:2015:a}. 
Viscoelastic estimates lump together resistance to deformation from both the solid and fluid constituents. 
The poroelastic shear modulus is the value for the drained solid, so will likely be somewhat lower than viscoelastic approximations, although a similar order of magnitude for most reasonably solid tissues.
The frequency dependence of both is interesting, since it means that there is a memory effect in the tissue dynamics that has not been accounted for in the equations of motion.
In order for this memory effect to be causal, in the sense that only past dynamics and not future dynamics affect the present state, certain constraints known as Kramers-Kronig relations must be satisfied by real and imaginary components of the moduli\cite{Sinkus:2007:a}.
There is still some controversy over whether these conditions are satisfied by moduli measured in vivo.
Finding a frequency dependence of the poroelastic parameters would put similar constraints on the equations of motion.

The introduction of $\beta_n$ in the poroelastic equations of motion creates a link between poroelastic and viscoelastic dynamics by means of the imaginary component of the parameter.
Past numerical studies attempted to bridge the connection between the two types of behavior\cite{McGarry:2015:a}.
A full analytical treatment is in the works - understanding where the dynamics are most similar and where they are most different is a way of gaining insight into the microstructure of tissue.
It also points towards better experimental design, since understanding the parameter spaces of the two can be harnessed to accentuate, and hence test for, behavior intrinsic to only one type of model.
Forthcoming work will focus on understanding this connection.\\

One might wonder what happens when we attempt to infer more properties.
Unfortunately, as the number of properties increases, the coarseness of the material meshes must also increase.
This is simply due to matching displacement constraints coming from MR measurements to the property degrees of freedom in the equations of motion.
Theoretical considerations aside, our numerical suite has a much harder time converging with the addition of an extra property to infer - a significant contraction of the domain of convergence being observed. 
We believe that this is a numerical issue, not a theoretical one, and an active research direction within our group.

As we move forward with in vitro and in vivo testing, our in silico results give us a good picture at how ignorant we can be when guessing our background values.
Hydraulic permeability reconstruction for noiseless data is robust to initial conditions ranging from an order of magnitude below, to $2+$ orders of magnitude above the ground truth background. 
Meanwhile, shear modulus alone has not been observed to diverge, albeit its convergence rate can get quite slow.
Adding noise to data introduces fluctuations that will affect reconstruction, as our group has shown in prior work \cite{Tan:2017:b}.
We found our results to be robust up to contrasts of an order of magnitude (it was small contrasts that were problematic) and believe that as long as signal to noise ratios are well above unity, noise should have only a minor effect on the domain of convergence.
\\

\subsection*{Concluding Thoughts}

Being able to image multiple poroelastic parameters will be a boon for preclinical diagnosis.
Getting our computational suite to work in vivo will allow us to apply poroelasticity to the brain, where both cerebro-spinal fluid and blood play an active roll in homeostasis. 
The robustness under a wide range of initial conditions at low frequencies bodes well for using intrinsic actuation via the cardiac cycle. 
Tumors tend to have shear moduli between one and two orders of magnitude higher than surrounding tissue - these will be as visible to poroelastic parameter reconstruction as they are for viscoelasticity.
The movement of fluid in the brain is critical in several venues.  
It is a mixture blood movement and interstitial fluid movement over the scale of a fraction of a voxel. 
We expect the dominant scale of fluid flow observed in MRE will be between that of perfusion which observes macroscopic bulk blood flow over a fraction of a meter and that of diffusion which observes microscopic fluid flow on a cellular scale. 
Observing fluid flow in this scale has potential to complement diffusion in characterizing edema and dementia.  
We postulate this scale of fluid flow will also illuminate tissue viability before diffusion becomes relevant allowing us to better identify and characterize reversible tissue damage.  
It might also provide additional information about functional changes secondary to brain activity.  \\

It is an exciting time to watch as MRE enters it's third decade of application.
As interdisciplinary as its roots are, it is no surprise that with maturity the field is digging deeper into the fertile soil of its disciplinary constituents.
Poroelasticity has been one of the crowning theoretical achievements of the geophysics community, and like every good idea, is ripe for cross-pollination.
Perhaps it is no surprise that a model developed to describe rocks and soils is finding use in describing living tissue when in 1510 Leonardo da Vinci penned\cite{DaVinci:1510:a}, into his {\it Codex Leicester}, that {\it
The Earth is a living body. Its flesh is the soil, its bones are the strata of rock, its cartilage is the tufa, its blood is the underground streams, the reservoir of blood around its heart is the ocean, the systole and diastole of the blood in the arteries and veins appear on the Earth as the rising and sinking of the oceans.}\\

\subsection*{Methods and Data Availability}
Simulations were performed using an in-house FORTRAN MRE Computational Platform written and developed over the past decade by group members.
Both forward and inverse problems were computed on the Discovery Cluster, a 3000+ core Linux cluster at Dartmouth College administered by Research Computing.
The data that support the findings of this study are available from the corresponding author upon reasonable request.\\

\subsection*{Acknowledgements}
This work was supported by the National Institute for Health Research Grant (No. $R01EB027577-02$). 
We would like to thank Research Computing at Dartmouth for their ever present ability with troubleshooting our cluster usage, and wish Susan Schwarz an adventurous retirement.\\

\bibliography{references.bib}
\bibstyle{unsrt}
\onecolumngrid
\section*{Figures}

\end{document}